\documentclass{aa}  

\usepackage{graphicx}
\usepackage{txfonts}
\usepackage{mhchem}
\usepackage{booktabs}
\usepackage{xcolor}
\usepackage{hyperref}

\begin{document}


    \title{Diffusive vs. non-diffusive paths to interstellar hydrogen peroxide}

   \subtitle{A machine learning-based molecular dynamics study}

   \author{J. Poštulka
          \inst{1}
          \and
          P. Slavíček
          \inst{1}
          \thanks{Corresponding authors: PS and GM.}
          \and
          J. K\"astner
          \inst{3}
          \and
          G. Molpeceres
          \inst{2}
          }

   \institute{Department of Physical Chemistry, University of Chemistry and Technology, Technick{\'a} 5, Prague 6, Czech Republic
        \email{petr.slavicek@vscht.cz}
         \and
        Departamento de Astrofísica Molecular, Instituto de Física Fundamental (IFF-CSIC), C/ Serrano 123, 121, 113bis, 28006 Madrid, Spain
        \email{german.molpeceres@iff.csic.es}
        \and
        Institute for Theoretical Chemistry, University of Stuttgart, Pfaffenwaldring 55, 70569 Stuttgart, Germany 
             }

   \date{Received \today; accepted \today}

 
  \abstract
   {Radical chemical reactions on cosmic dust grains play a crucial role in forming various chemical species. Among different radicals, the hydroxyl (OH) is one of the most important ones, with a rather specific chemistry.}
   {The goal of this work is to simulate the recombination dynamics of hydroxyl radicals and the subsequent formation of hydrogen peroxide (\ce{H2O2}).}
   {We employed neural network potentials trained on ONIOM(QM/QM) data, combining multi-reference (CASPT2) and density functional theory (DFT) calculations. This approach allows us to model the recombination of hydroxyl radicals on ice surfaces with high computational efficiency and accuracy.}
   {Our simulations reveal that the initial position of the radicals plays a decisive role in determining recombination probability. We found that the formation of hydrogen-bond between radicals, competes with the formation of hydrogen peroxide, reducing the recombination efficiency, contrary to expectance. This competition reduces the recombination probability for radicals that are initially formed approximately 3 \AA{} apart. Recombination probabilities also depend on the kinetic energy of the added radicals, with values around 0.33 for thermal radicals and a wide range of values between 0.33 and 1.00 for suprathermal OH radicals.}
   {Based on our calculations we provide recommendations for introducing OH radical recombination into kinetic astrochemical models, differentiating between thermal and suprathermal radicals. The recombination behavior varies significantly between these two cases: while thermal radicals are sometimes trapped in hydrogen-bonded minima, 
   the case of suprathermal radicals varies with the added energy. Our most important conclusion is that OH radical recombination probability cannot be assumed 1.0 for a wide variety of cases.}


   \maketitle
%

\section{Introduction} \label{sec:introduction}

The formation of interstellar water, actually amorphous solid water (ASW), is a subject that has sparked a lot of interest over the years, \citep[see, for example][]{VanDishoeck2013}. While our understanding of the water formation process has evolved, the conclusions obtained in the early models \citep{tielens_model_1982} are consistent with the current experimental \citep{Ioppolo2008, Oba2009, romanzin_water_2011, dulieu_experimental_2010, Ioppolo2010, Lamberts2014} and theoretical understanding \citep{Cuppen2010, lamberts_water_2013, Lamberts2014, Meisner2017, Molpeceres2019}. Essentially, there is a large body of evidence in that ASW forms atop interstellar dust grain through reactions where the hydroxyl radical (OH) takes a protagonist role through the O-atom reaction network \citep[see, for example, Figure 17 of][]{VanDishoeck2013}; namely:
\begin{align}
    &\ce{O ->[H] \textbf{OH} ->[H] H2O } \label{eq:o_hyd} \\
   &\ce{H2 + \textbf{OH} -> H2O} \\
    &\ce{O3 + H -> \textbf{OH} + O2 ->[H] H2O } \\
    &\ce{OH + \textbf{OH} -> H2O + O} \label{eq:oh_oh_triplet} \\
    &\ce{OH + \textbf{OH} -> H2O2 ->[H] H2O + H2 }. \label{eq:oh_oh_singlet}
\end{align}
In addition to the above, the chemistry of \ce{O2} also partakes in the formation of \ce{H2O} without the need of OH radicals:
\begin{equation}
    \ce{O2 ->[H] HO2 ->[H] H2O2 ->[H] H2O + H2}.
\end{equation}
In diffuse and translucent clouds, or in photo-processed regions in a broad sense, water ice is subjected to energetic chemistry \citep[see, for example][]{Oberg2016} where the main dissociation channel is:
\begin{equation} \label{eq:dissociation}
    \ce{H2O + $h\nu$ -> \textbf{OH} + H},
\end{equation}
leading again to the OH radical.

In view of what was presented above, the immediate conclusion is that ASW in photo-processed regions is constantly formed and destroyed, with OH radicals present in every reactive or energetic event. Considering that \ce{H2O} ice is ubiquitously detected in infrared observations of interstellar ices \citep{Oberg2011, Boogert2015, mcclure_ice_2023, dartois_spectroscopic_2024}, it is implied that either \ce{H2O} formation rates surpass destruction ones or formation and destruction are in a pseudoequilibrium. The most obvious ``reconstruction'' pathway involves the \ce{OH + H -> H2O} recombination (rightmost reaction in Equation \ref{eq:o_hyd}). However, it is expected that the nascent H atom will either desorb or diffuse farther from the reaction partner due to the excess energy of the fragmentation, which greatly surpasses the meager binding energy of H on \ce{H2O} \citep{Hama2012}. An additional reconstruction pathway was studied by \citet{redondo_reconstruction_2020} recently. In their study, \citet{redondo_reconstruction_2020} showed the viability of Reaction \ref{eq:oh_oh_triplet} using theoretical methods and finding the process viable. This study was performed on the triplet potential energy surface, meaning that both OH radicals have parallel spins at the time of encountering. The same reaction in the singlet state produces, as a byproduct $^1$O, an excited state that questions whether such a route has low or no barriers and is exothermic. In fact, the expected product of the 2 OH recombination in the singlet state (leftmost reaction in Equation \ref{eq:oh_oh_singlet}) is the production of hydrogen peroxide (\ce{H2O2}) that is directly linked with \ce{H2O} via a H-abstraction reaction \citep{Lamberts2016}. We present Reaction \ref{eq:oh_oh_singlet} again because it is the chemical process that we simulate in this work:
\begin{equation}
    \ce{\textbf{OH} + \textbf{OH} -> H2O2} \label{eq:central}.
\end{equation}
Both triplet and singlet channels for the \ce{OH + OH} reaction are possible in the UV-irradiated icy dust grain, with the triplet one being statistically favored. While the triplet channel is easily investigated owing to the kinetic barriers associated with this route, the singlet channel is less straightforward as it is assumed to readily produce \ce{H2O2} upon encounter with a 1.00 branching ratio. Nevertheless, the strong interaction between \ce{OH} and \ce{H2O} \citep{miyazaki_photostimulated_2020,OH-Watanabe} can actually lower this branching ratio, making 2 \ce{OH} adsorption favorable over reaction for certain OH-OH orientations, in a similar flavor to what was found for a wide collection of radicals on ASW \citep{enrique-romero_quantum_2022}.

Neutral-neutral two-body reactions, a term encompassing the aforementioned radical-radical recombinations, can mechanistically proceed in several ways in interstellar dust grains. Without any input of external energy, i.e. under strictly thermal conditions, the rate of the reaction depends on the diffusion of the participating reactants in addition to the activation barriers or lack thereof. Therefore, adsorbates that are not mobile on the surface will hardly react, leaving only a handful of radicals and molecules, e.g., the H, N, and O atoms, or CO molecules \citep{Hama2012, Minissale2016a, furuya_diffusion_2022} to diffuse and react at 10 K. However, the presence of complex organic molecules (COMs) not explainable by reactions with the above-mentioned species in very cold prestellar cores \citep{Bacman2012, Jimenez_Serra_2016, scibelli_prevalence_2020} used to constitute a challenge for astrochemical models. Energetic processes, either initiated by cosmic rays \citep{shingledecker_cosmic-ray-driven_2018}, photons \citep{mullikin_new_2021} or chemical reactions \citep{Jin2020} have emerged as viable alternatives to explain the mobility of otherwise immobile species, at least in \ce{H2O} ice, \citep[see][for results on other ices]{molpeceres_enhanced_2024}. Considering the main dissociation channel of \ce{H2O} (Reaction \ref{eq:dissociation}) by photons and cosmic rays, in addition to the prevalence of OH radicals in hydrogenation reactions, e.g. \ce{O + H -> OH} makes OH radicals arguably the most abundant radical produced by energetic mechanisms. In fact, with a null mobility at 10 K \citep{miyazaki_direct_2022} \citet{Ishibashi2021} showed experimentally that suprathermal OH radicals are able to react on an ASW surface. 

In this paper, we aim to deepen our knowledge of the chemistry of suprathermal OH radicals by focusing on a detailed study of Reaction \ref{eq:central} using molecular dynamics simulations driven by machine-learned interatomic potentials (MLIP).\footnote{A possible competitive reaction could be \ce{OH + OH -> H2O + O} in the singlet channel. However, this reaction is endothermic in the gas phase by 1.5 eV at the ground level of the theory considered in this work (CASPT2(6,4)/6-31+G(d,p)), Section \ref{sec:methods}.} Following our previous study for the hydrogenation of the phosphorus atom \citep{molpeceres_reaction_2023}, here we increase the complexity of our system and study the formation of a polyatomic molecule (\ce{H2O2}) from the recombination of two OH radicals. We focus on the singlet potential energy surface (PES) that leads to the molecule of interest. From an astrochemical point of view, our goal is to enhance the description of energetic chemistry in the early stages of the star formation cycle, with a strong focus on the reformation of the ASW ice mantle after interaction with energetic photons. 

The techniques of \textit{ab initio} simulations are becoming widely used in astrochemistry, as they provide insight into processes that are challenging to access experimentally. \citep{astrochem_methods} Molecular dynamics simulations are typically conducted using density functional theory (DFT) as the computational framework, with key stationary points often benchmarked using coupled cluster methods for higher accuracy. However, single-reference methods like DFT and coupled clusters can encounter significant limitations when orbital degeneracies arise, such as in the modeling of excited states, multiple radicals, or homolytic bond cleavage. \citep{Cramer} In such cases, multi-reference methods are essential. Among these, the Complete Active Space Self-Consistent Field (CASSCF) method is widely employed, as it effectively captures static electron correlation. To achieve quantitative accuracy, however, the inclusion of dynamic correlation is critical. In this work, we address this by employing a perturbation correction to CASSCF, leading to the Complete Active Space with Second-Order Perturbation Theory (CASPT2) approach.  

Our paper is structured as follows. In Section \ref{sec:methods} we introduce our theoretical and structural models for the MLIP and surface slabs. Section \ref{sec:results} is dedicated to an exhaustive description of all the factors influencing Reaction \ref{eq:central} on ASW. After such a theoretical description, our results are put in an astrophysical context in Section \ref{sec:discussion}, including a summary of our findings, recommendations to include these results in astrochemical models, as well as the impact of our findings in our current understanding of interstellar \ce{H2O} formation and suprathermal chemistry. We conclude with a brief summary of the work, presented in Section \ref{sec:discussion:conclusions}.

\section{Methodology} \label{sec:methods}

\begin{figure*}
  \centering
  \includegraphics[width=0.75\linewidth]{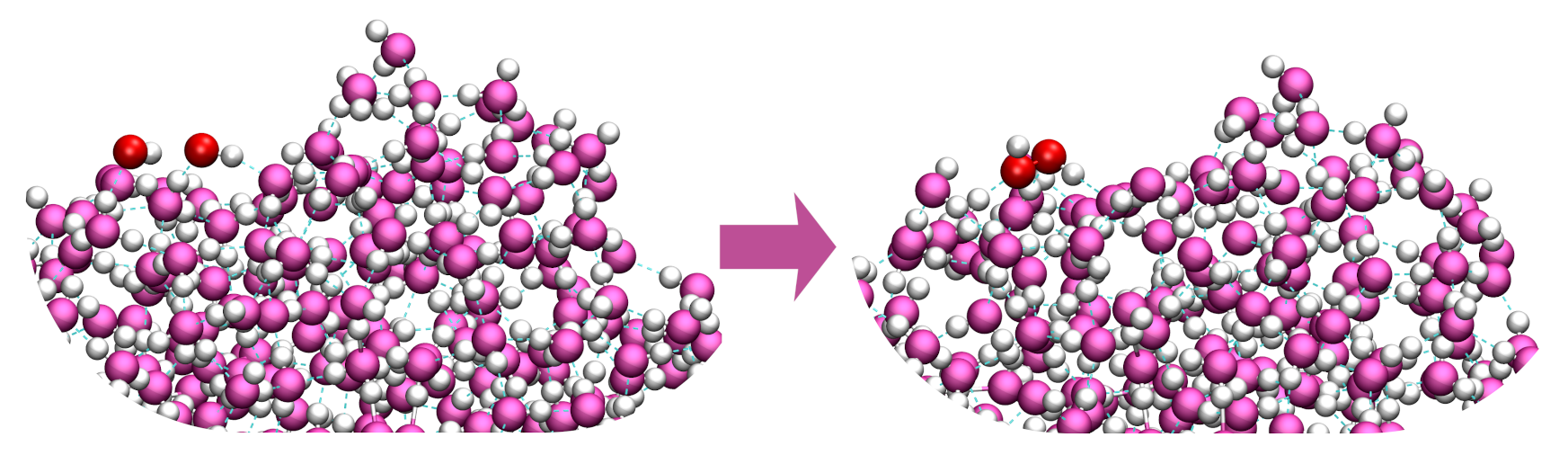}
  \caption{Initial and final geometry of one recombination trajectory, unbound radicals on the left and hydrogen peroxide on the right.}
  \label{fig:snapshot}
\end{figure*}

We have investigated the recombination dynamics of two hydroxyl radicals on a water surface. Such a task requires long simulations (from tens to hundreds of picoseconds) and a large system to cover both the reacting radicals and the surface of water ice. For that, we have constructed a reactive machine learning force field, enabling us to simulate (maximum) nanosecond trajectories with hundreds of molecules.

\subsection{Construction of the training set: Description of the ground truth method for our MLIP potential}

\begin{figure}
  \centering
  \includegraphics[width=0.75\linewidth]{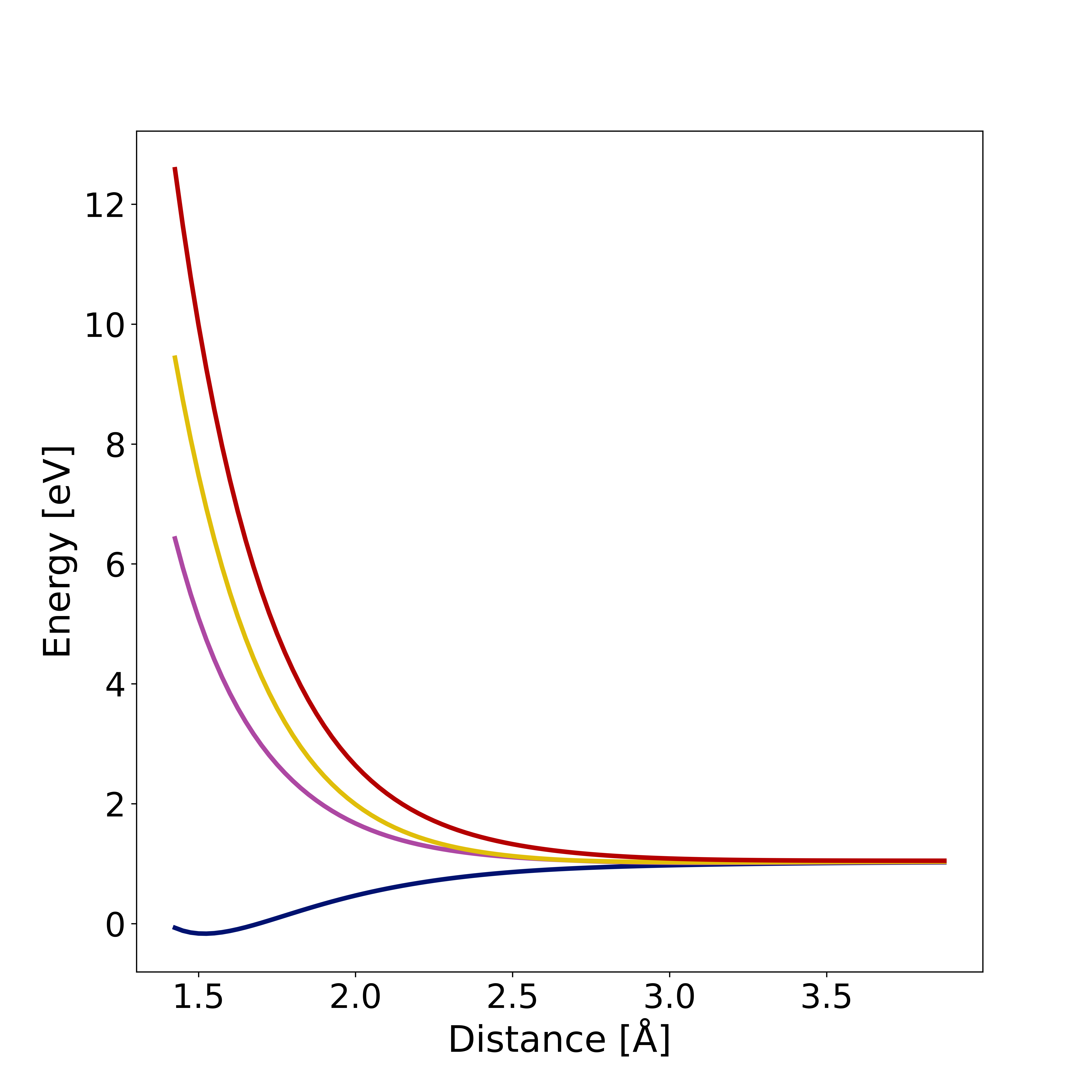}
  \caption{Rigid scan of potential energy surfaces of all 4 states included in the CASPT2 calculations.}
  \label{fig:AS}
\end{figure}

Most work when producing MLIP is to choose and label training set data points. Data set structures in our case were mainly 20-30 molecule clusters of water with two hydroxyl radicals, and less than 10\% of the set were geometries of pure water clusters (see Table \ref{tab:1}). As a reference method to calculate energies and forces of the data set, a combination of state-averaged complete active space (SA4-CASPT2(6,4)/6-31+G(d,p)) \citep{CASPT2,CASPT2_2} averaging four lowest singlet electronic states and Perdew-Burke-Ernzerhof functional (PBE(D3BJ)/def2-SVP) \citep{PBE,D3} was used. The active space in the CASPT2 calculations constituting from 6 electrons and one virtual orbital proved to be the best choice considering the smoothness of the potential energy surface along the inter-radical distance, which was the key parameter for our calculations. Smaller or slightly larger active spaces showed discontinuities in the rigid scans of PES along the inter-radical distance, which made the 6,4 active space the optimal choice, as shown in Figure \ref{fig:AS}. The CASPT2 calculations were performed using Molpro 2012.1.11 package \citep{MOLPRO_1,MOLPRO_2,MOLPRO_3} and the DFT calculations were done using the ORCA 5.0.3 code \citep{ORCA}. In the training set, both OH radicals were described by the multi-reference CASPT2 method. Water molecules were described at the PBE level because calculating the whole system at the CASPT2 level would be computationally intractable. Both theoretical treatments were combined in a QM/QM manner using subtractive ONIOM (Our own N-layered Integrated molecular Orbital and Molecular mechanics) embedding scheme \citep{ONIOM} with the electrostatic effects of the outer region on the CASPT2 core region wavefunction treated by adding CHELPG (CHarges from ELectrostatic Potentials using a Grid-based method) charges \citep{CHELPG} of water atoms as point charges in the CASPT2 calculation. Energies and gradients from both the multi-reference and the DFT calculations were then subtracted using our own script, completing the simplest ONIOM scheme. This approach combines computational feasibility with high accuracy, putting the focus on a good description of the radical-radical interactions. It is important to indicate that the DFT part of the QM/QM embedding was simulated using a broken symmetry wavefunction.

A sampling of the configuration space for the MLIP training set was done using several exploratory molecular dynamics simulations, similarly as in our previous works \citep{Molpeceres2020, Molpeceres2021a, molpeceres_reaction_2023}. Sampling simulations in the canonical ensemble (NVT, i.e., keeping the number of particles, volume, and system temperature constant) were performed at 300 K, using Nosé--Hoover chains of 4 thermostats, to sample interactions of the OH radical with the water surface. Molecular dynamics was simulated using in-house code ABIN \citep{ABIN}. Because hydroxyl radicals bind strongly with \ce{H2O} and their actual diffusion, even at 300 K, is beyond our simulation timescales, we performed steering dynamics interfacing the ABIN code \citep{ABIN} with the PLUMED library \citep{PLUMED_1,PLUMED_2,PLUMED_3}, ensuring a sufficient configurational sampling. Additionally, remaining non-sampled spots in the OH-OH distance interval were then recovered via an \textit{a posteriori} umbrella sampling \citep{umbrella}. After obtaining all the training points, we randomly divided them into 75\% for training, 10\% for the validation set and 15\% for the test set in our MLIP protocol.

\begin{table}[t]
\begin{center}
\caption{Types and quantity of clusters in the training set.}
\label{tab:1}
\begin{tabular}{cc}
\toprule
Cluster type & Number of data points \\
\bottomrule
10 H$_{2}$O & 450 \\
20 H$_{2}$O & 693 \\
30 H$_{2}$O & 672 \\
70 H$_{2}$O & 1044 \\
2 OH and 20 H$_{2}$O   & 33963 \\
2 OH and 30 H$_{2}$O   &  2383 \\
\bottomrule
\end{tabular}
\tablefoot{Temperature for the sampling is always 300 K and the simulations are performed in the NVT ensemble. Some points are extracted from enhanced sampling dynamics (e.g. steering dynamics of umbrella sampling), see text.}
\end{center}
\end{table}

\subsection{Training of the model}

\begin{figure}
  \centering
  \includegraphics[width=0.75\linewidth]{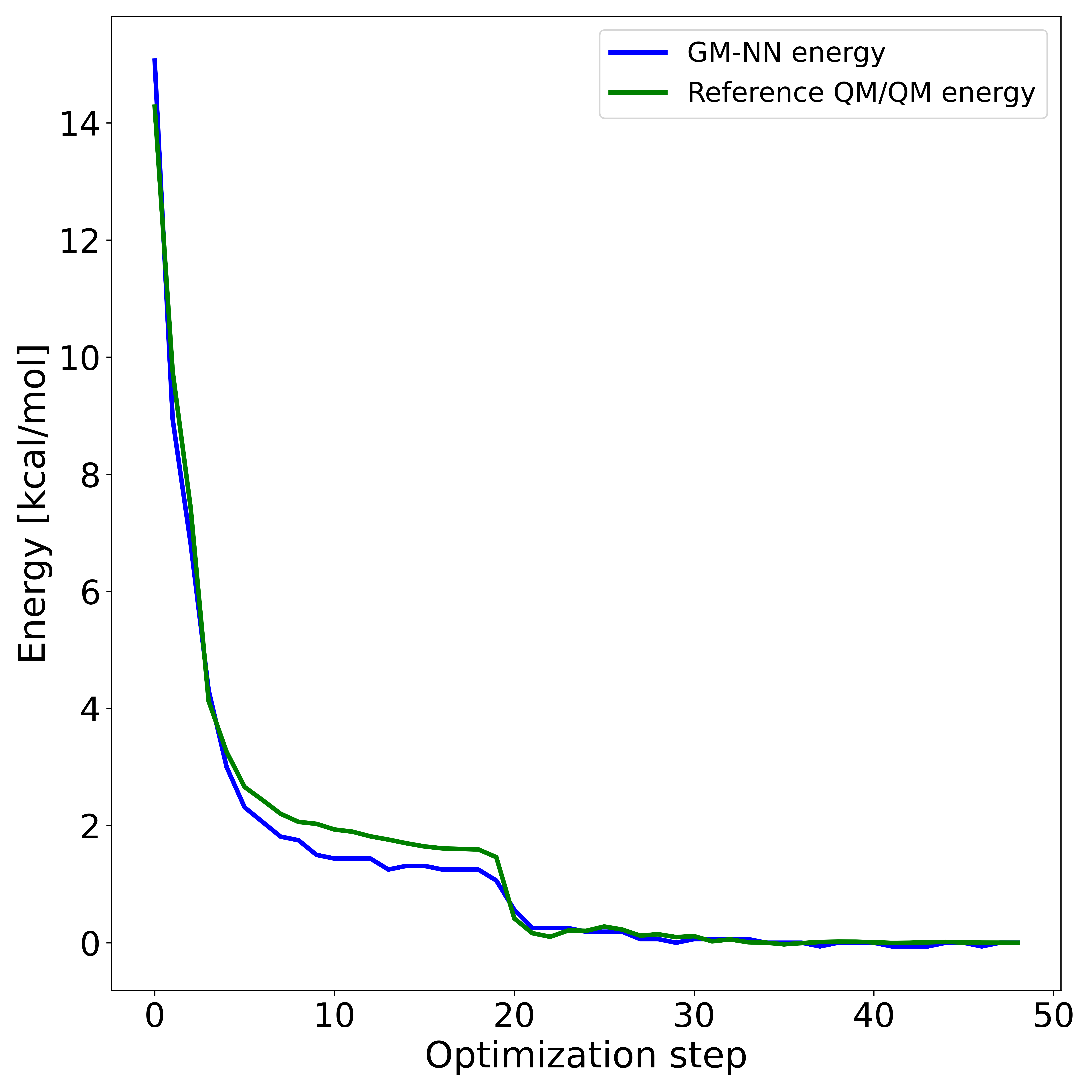}
  \caption{Performance of the model in structure optimization calculation compared against reference QM/QM method. }
  \label{fig:opt}
\end{figure}

\begin{figure}
  \centering
  \includegraphics[width=0.90\linewidth]{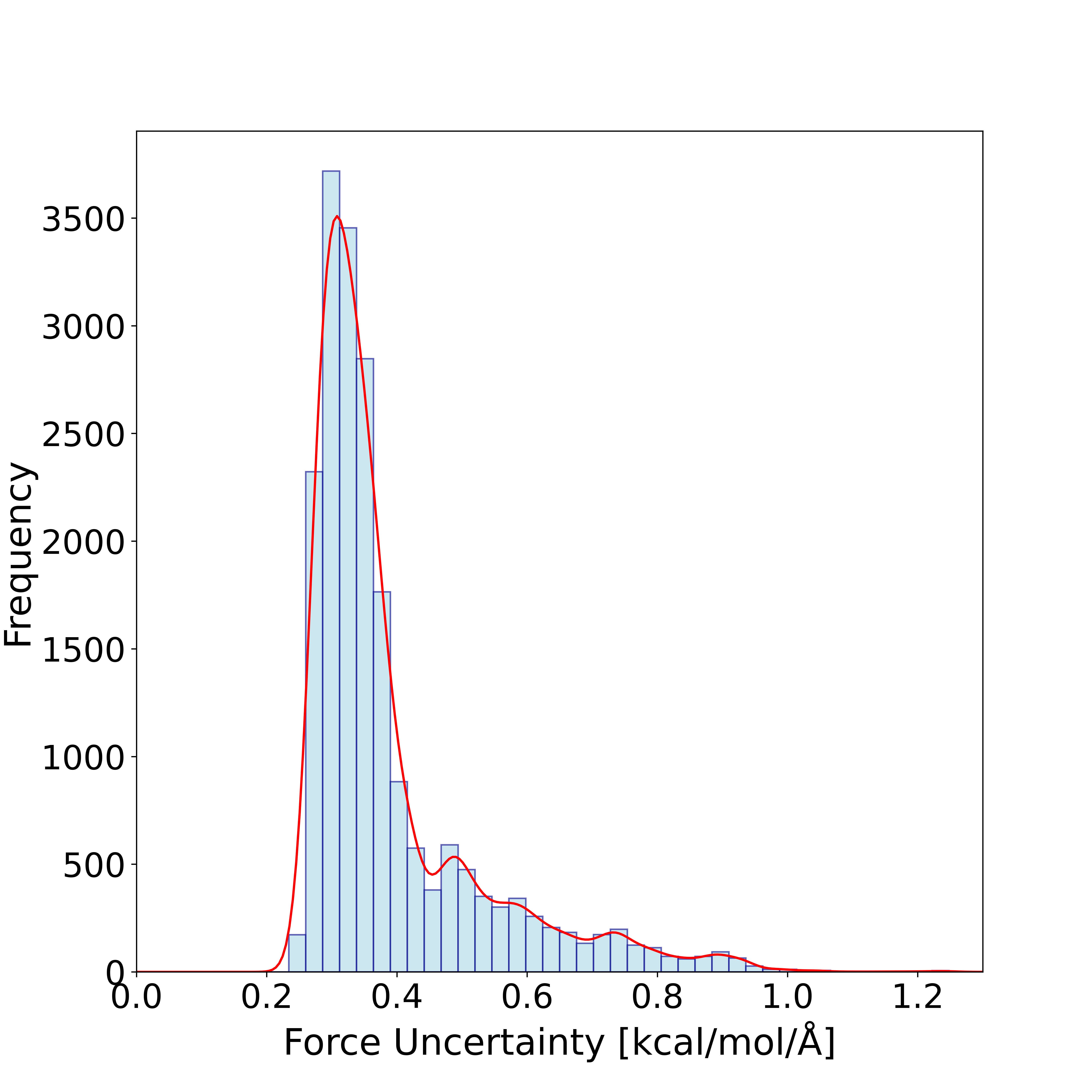}
  \caption{Force uncertainty tracked along the trajectory of MD steering dynamics to determine the performance along the reaction coordinate. Uncertainty was computed from the ensemble of 3 models and was tracked during the whole time of the simulation.}
  \label{fig:steering_force}
\end{figure}

The MLIP was obtained using the Gaussian-Moment Neural Network (GMNN) code \citep{gmnn,Zaverkin2021}. A neural network with 2 hidden layers and 512 nodes each was trained on a data set containing 39,205 geometries, electronic energies and atomic forces. Three independent MLIPs were trained at the same time on 75 \% of the data set in the training set, 10 \% in validation. Training was done for 1500 epochs on Nvidia GeForce RTX 3090 GPU. A cutoff radius of 5.5 \AA{} was used for Gaussian moments' descriptor. The usage of an ensemble of models allows us to determine the uncertainty of the predicted forces during our simulation from the disagreement of trained models. We achieved mean absolute error in forces under 0.5 kcal mol$^{-1}$ \AA$^{-1}$ for the training and validation set as well as for the test set, which was previously unknown to the neural network. 

Apart from evaluation on a predefined test set, we performed an optimization test. The calculation was done on a randomly sampled cluster of 20 molecules to be able to compute it efficiently with not only our model but also the reference method. In Figure \ref{fig:opt} an energy curve during optimization is presented to show how well the model performs in this manner. The predicted energy tracks the reference optimization curve very well, with only minor deviations from the reference values. This test again shows good performance in the prediction of the shape of the reference potential energy surface. 

Finally, we conducted in-depth testing of the inter-radical potential, giving our main goal was to study recombination of radicals. In order to test the interaction of radicals along the reaction coordinate, we used steering molecular dynamics. The simulation started with \ce{H2O2} formed on a cluster surface, and a moving artificial potential was added to separate oxygen atoms and form radicals at the end of the simulation. This way, the radicals were sampled at distances between the bond length of about 1.5~\AA{} and a potential cutoff radius of 5.5~\AA. In Figure \ref{fig:steering_force} the force uncertainty by the MLIP in the trajectory consistenly stays below the 1 kcal/mol/\AA{} threshold. Such performance exceeded our expectations, and ensures the accuracy of our potential. Further validation tests of the potential are gathered in Appendix \ref{sec:apA}).

\subsection{Creation of surface slabs}

Our simulations are run on the surface of a spherical-like amorphous water cluster. We prepared a cluster of 500 water molecules with 2 OH radicals on the surface. We began the construction of the cluster by randomly positioning 500 water molecules in a spherical volume of approximately 1.00 g cm$^{-3}$ density. From this cluster, in the first place, we ran a molecular dynamics trajectory at 300 K for 500 ps to generate different cluster conformations. The geometry of the system was sampled every 250 fs after an equilibration period of 50 ps. For every sampled water cluster, we deposited OH radicals on its surface by randomly positioning them in the proximity of the water cluster. The distance between the two radicals was preset to a specific value in the interval from 2.0 to 5.5 \AA{} (see Section \ref{sec:results}), being the inter-radical distance one of the sampled magnitudes in our study (see Section \ref{sec:results}). Before the production run of molecular dynamics, clusters were thermalized at 10 K for the duration of 20 ps. The distance between radicals was constrained for the whole time in order to keep the randomized sampling of the inter-radical distance between 2.0 and 5.5 \AA.

\subsection{Calculating potential of mean force}

To determine the range of attractive forces between the radicals, understand the dynamics of the system, and estimate ideal conditions for our production calculations, we performed a long simulation using the metadynamics approach \citep{MetaD1,MetaD}. For this purpose a long trajectory was run with a time step of 0.5 fs and a total simulation time of 2.5 ns with the starting positions corresponding to already formed \ce{H2O2} on the surface of a water cluster containing 500 molecules. We were adding bias potentials to the distance of oxygen atoms. The potential was added every 1000 steps (500 fs) in the form of a Gaussian function with a height of 1 kJ mol$^{-1}$ and width of 0.1 \AA. An additional strong wall potential was placed at a value of 6.0 \AA{}  of the distance of radical (just slightly higher than our training cutoff (see above)), to prevent them from separating further. Setup of this simulation was performed with PLUMED program \citep{PLUMED_1,PLUMED_2,PLUMED_3} which was also used to reconstruct the free energy surface from the simulation at a temperature of 10 K.

\subsection{Generation of initial conditions for production trajectories} \label{sec:methods:recomb_traj}

We performed a set of MD trajectories to study the recombination of hydroxyl radicals on a water surface using the MLIP. For every geometry from the initial equilibrated position, see above, a total of 5 trajectories were performed with different momenta given to the OH radicals, one with initial velocities extracted from the previous NVT equilibration at 10 K and another 4 trajectories associated with the different extra momenta added to one of the hydroxyl radicals to mimic the production of suprathermal OH radical by different mechanisms, for example through chemical reactions or photodissociation of a \ce{H2O} molecule. The list of added energies is \{0.1, 0.3, 0.5, and 1.0\} eV. The energy was added by appropriate scaling of the velocity vector from the initial velocities. To prevent radical evaporation by this added energy, only the projection of the velocity vector to the surface tangent to the cluster surface was prolonged. Trajectories were simulated for 500 ps or preemptively terminated after a recombination occurred (to save computational resources). These simulations were performed in the NVE regime, so the reactivity would not be influenced by the artificial thermostat. Molecular dynamics using the MLIP are performed using the Atomic Simulation Environment (ASE) \citep{ASE_1,ASE_2}.

\subsection{Motion of hydroxyl radical}

Finally, simulations similar to those shown in Section \ref{sec:methods:recomb_traj} (i.e, NVE on a 10 K water surface) were performed to study the motion of only a single OH radical. In this case, apart from simulations with added kinetic energy (of 0.1, 0.3, 0.5 or 1.0 eV), we performed also a single NVT simulation with a constant temperature of 10 K was performed using ASE environment, serving the latter as a blank (no mobility, see Section \ref{sec:motion}). The goal of such simulation was to observe the motion of OH without the influence of a second radical and possible recombination.


%
\section{Results} \label{sec:results}

The main purpose of this work is to determine recombination probabilities for Reaction \ref{eq:central} with respect to the initial position of the two radicals, and their initial kinetic energy. Such information allows us to discuss the influence of diffusive processes versus the influence of formation routes of the hydroxyl radical. After introducing the energy surface in the first section of the results, we focus on the dynamics of the reaction \ce{OH + OH -> H2O2} (Reaction \ref{eq:central}) in the second, finally we investigate the motion of OH radical on the surface of water.

\subsection{Potential of mean force}

\begin{figure}
  \centering
  \includegraphics[width=0.75\linewidth]{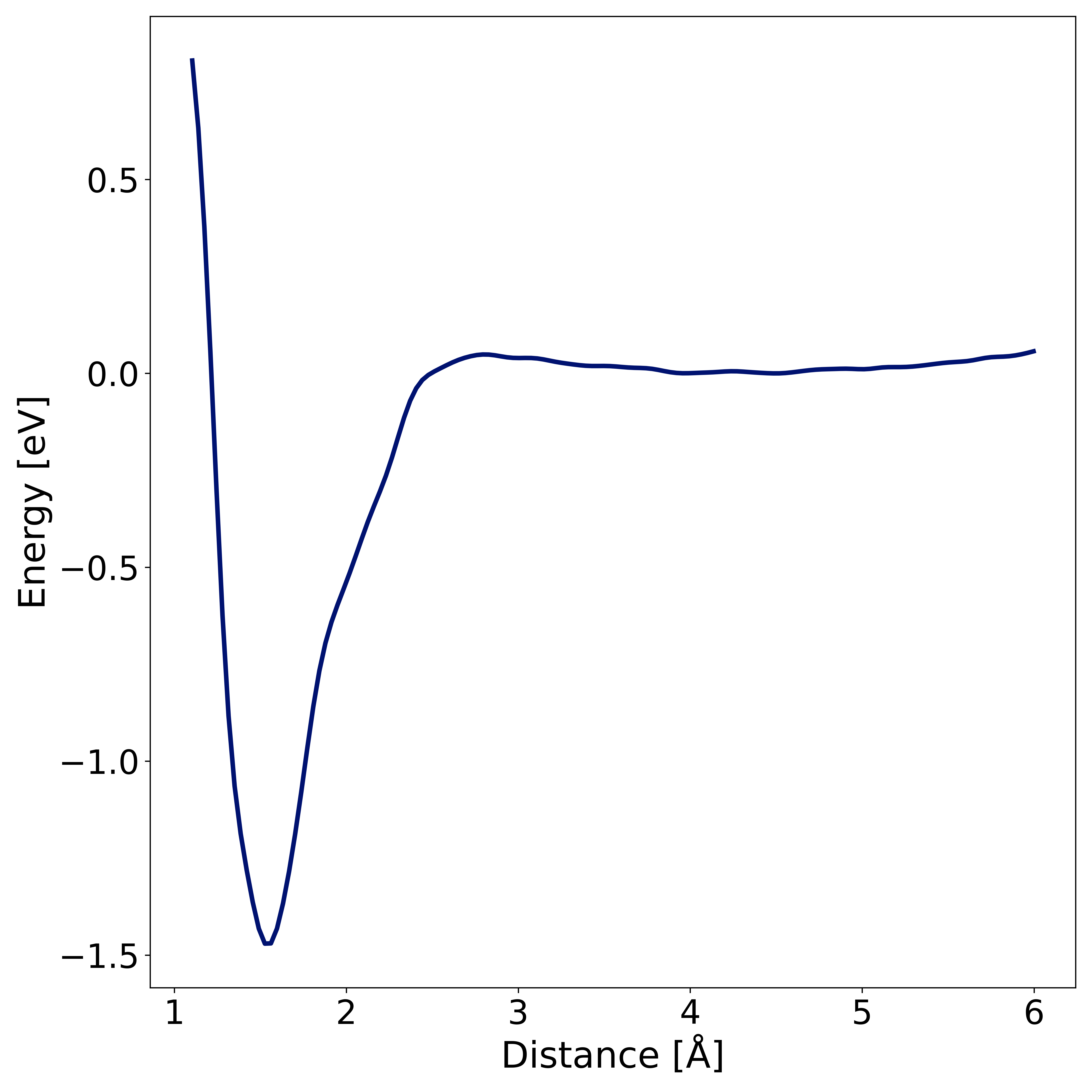}
  \caption{Potential of mean force for inter-radical distance, showing the free energy curve obtained from the metadynamics simulation.}
  \label{fig:PMF}
\end{figure}

To determine the range and shape of the inter-radical potential for OH radicals, a free energy surface (FES) was reconstructed from bias potentials added in a long metadynamics simulation. The free energy curve for inter-radical distance is presented in Figure \ref{fig:PMF}. The inter-radical potential is very attractive up to a distance of 2.5 \AA{} where the FES curve quickly flattens and changes only minimally for greater distances. It ought to be mentioned, that full convergence was not achieved, given the size of the system and therefore very large conformational space. With that in mind, we look at this figure in a qualitative manner. Nevertheless, we can convincingly say that radicals have to approach each other to a distance closer than 2.5 \AA{} in order to be trapped in the deepest potential well, the \ce{H2O2} molecule. The well depth of the potential does not allow radicals to dissociate without additional energy provided by photoexcitation or similar processes. We can observe a small barrier at the distance of 2.8 \AA{} with a height of 0.05 eV, equivalent to roughly the thermal energy of 600 K. The nature of this barrier is difficult to identify, as it is somewhat low to diffusion barriers arising from the \ce{HO-H2O} H-bond breaking with the surface, and associated with barriers of around 1600 K \citet{miyazaki_direct_2022}. Another possibility is that is a barrier from the OH...OH complex that we found in our dynamical calculations (See section \ref{sec:results:recombination}). The low inter-radical distances make us think that the latter is a more likely explanation. In either case, it is important to clarify what the FES represents. Our FES is a reconstruction over a limited time (2.5 ns) of all possible events occurring in that timelapse in a metadynamics run. In other words, the FES shows a balance between the barrierless reaction and the process with a barrier. Because the OH...OH complex will, in many cases, lead to a barrierless recombination, the contribution of the activated processes in the FES is weighted down when barrierless paths are found. We emphasize, however, that finding a barrier in the FES does not mean that Reaction \ref{eq:central} has an intrinsic barrier by itself, but rather that some reactive orientations on an ASW do have a barrier, depending on the position on the ASW.

We can see the curve rising at the largest distance of 6~\AA{} again. This is, however, just an artifact of the repulsive wall, which was placed at 6.5 \AA{} to prevent the radicals from separating further in the metadynamics simulation. The potential should be basically flat for larger distances, where the influence of the second radical disappears.

\subsection{Recombination reaction} \label{sec:results:recombination}

\subsubsection{Influence of the initial position}

\begin{figure*}
  \centering
  \includegraphics[width=0.75\linewidth]{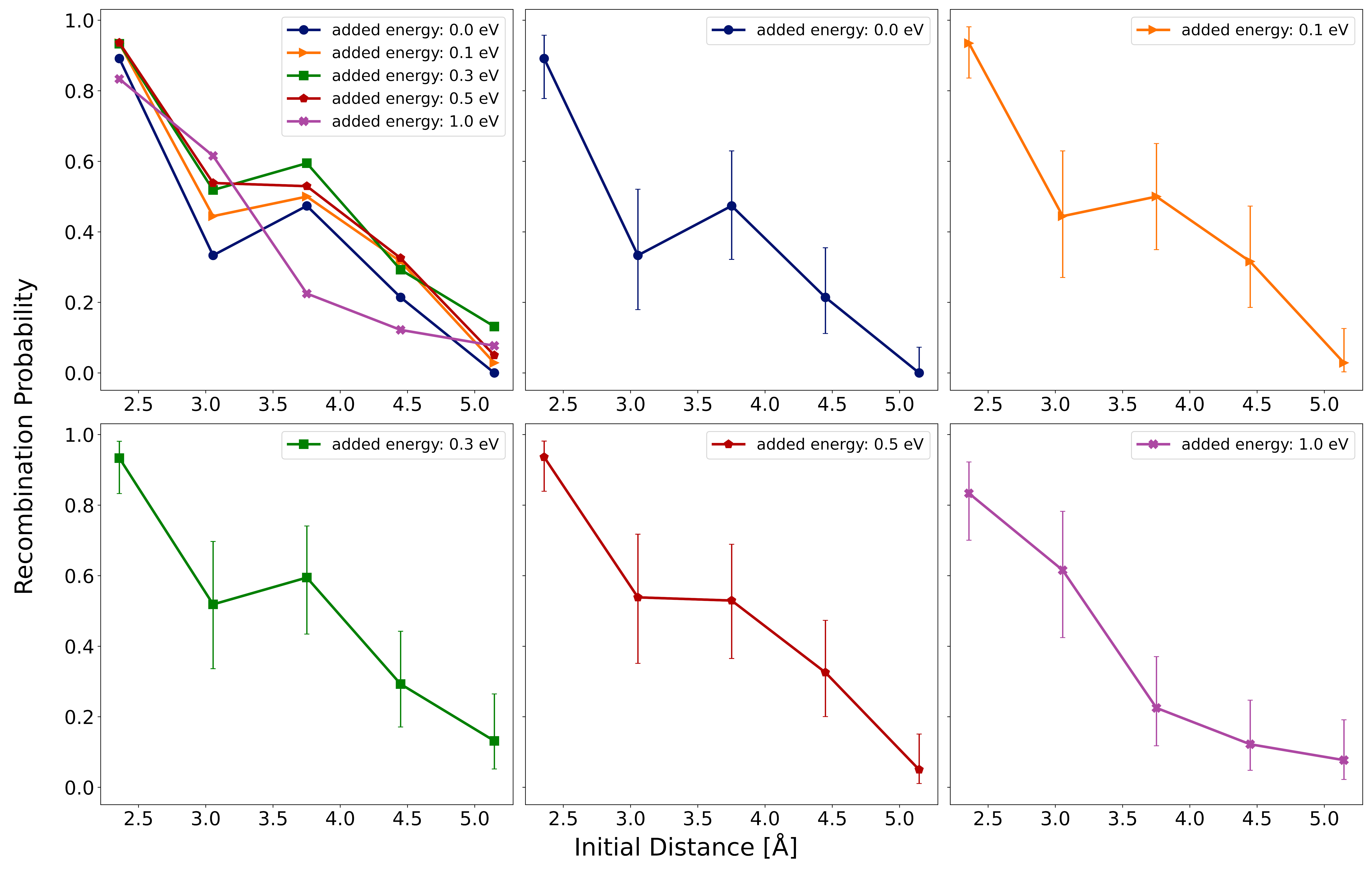}
  \caption{Recombination probability as a function of the initial distance of the OH radicals. The first top-left plot presents the results for all the energies combined. The following graphs present individually the datasets for different added energy, with the error bars included in the plots. Error bars represent Jeffreys Bayesian Interval generated by python library Statsmodels \citep{statsmodels}.}
  \label{fig:recomb_distance}
\end{figure*}

\begin{figure}
  \centering
  \includegraphics[width=0.75\linewidth]{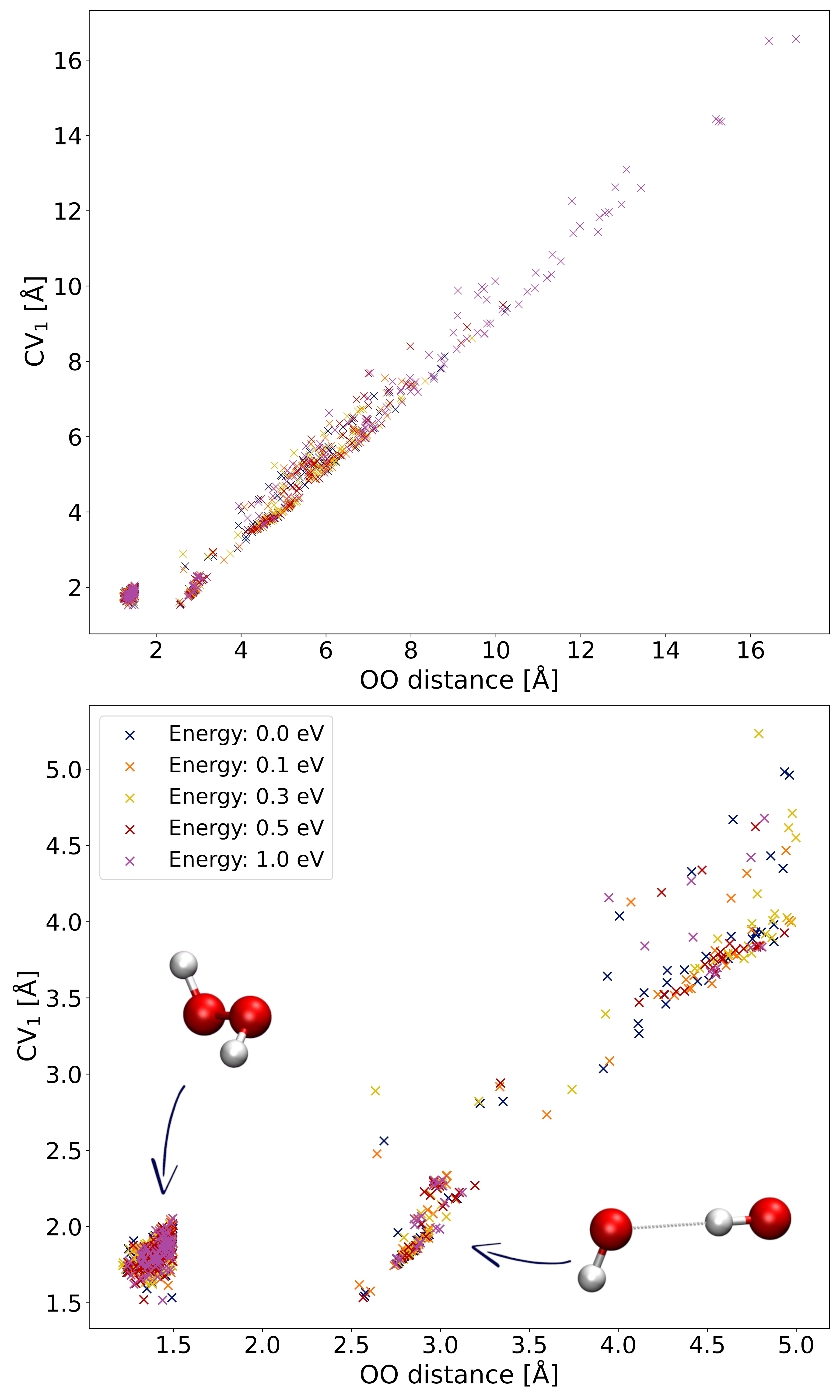}
  \caption{Plot of final geometries of all trajectories, the upper plot shows all endpoints of trajectories in the 2D graph of $\textrm{CV}_{1}$ and O-O distance, the lower plot presents data from the same trajectories zoomed on H-bonding and hydrogen peroxide conformers.}
  \label{fig:2Dplot}
\end{figure}

We investigated the OH recombination probability as a function of the initial distance of radicals. The results are shown in Figure \ref{fig:recomb_distance}. In the top left panel, we see that the recombination probability in general decreases with growing initial distance, thus supporting the expected outcome of the close radicals having a higher chance to recombine. Even at the closest distance of 2.5 \AA{} there is a finite probability for the radicals to avoid recombination, due to an unfavorable direction of initial velocities. With higher added kinetic energies, it is reasonable to assume that some trajectories will leave the reactive potential because the added energy is sufficient to overcome the attractive potential. This is especially true when the velocity vector points opposite to the direction of recombination. However, even when no kinetic energy is added to either of the two OH radicals, we observe some trajectories where there is no recombination. This seems counterintuitive at first, since the energy gradient is supposed to be high and pointing towards the minimum energy \ce{H2O2} situation. A careful inspection of these few outliers reveals the presence of hydrogen-bonded OH...OH strong pairs that are key in the recombination efficiencies and dynamics. These situations are central in our discussion and will be investigated in more detail later in the text. Here it suffices to indicate that the presence of these complexes at distances $\sim$2.5 \AA{} is rare (Figure \ref{fig:2Dplot}) which correlates with the high (yet not unity) recombination probability at low distances. No dissociation of \ce{H2O2} ever occurs, once the molecule is formed, which was expected considering the depth of the minimum listed in Table \ref{tab:2}.

Continuing with other starting energies, for the case of no additional energy, we observe no recombination above 5 \AA, within the simulation time. Recombination takes place when radicals are trapped in the inter-radical attractive potential, radicals either have to approach one another by diffusion or at a smaller distance they slip to the preferred binding site on the surface, when being initially in an unfavorable position. This line of thinking leads inevitably to the assumption that the closer the radicals, the higher the probability of recombination. However, if we look closely on the separate plots of the trajectories with different kinetic energy added to the radicals, we observe a minimum in the probability at around 3{~\AA} for the cases where 0, 0.1  and 0.3 eV of kinetic energy were added. This minimum seems counter-intuitive as we would expect that the closer the radicals are, the higher the probability of recombination. As we noted above if we suppose a diffusive nature of this process it should lead to a continual decrease with energy, thus this minimum suggests a more complex mechanism of the whole process. 

\begin{figure}
  \centering
  \includegraphics[width=0.6\linewidth]{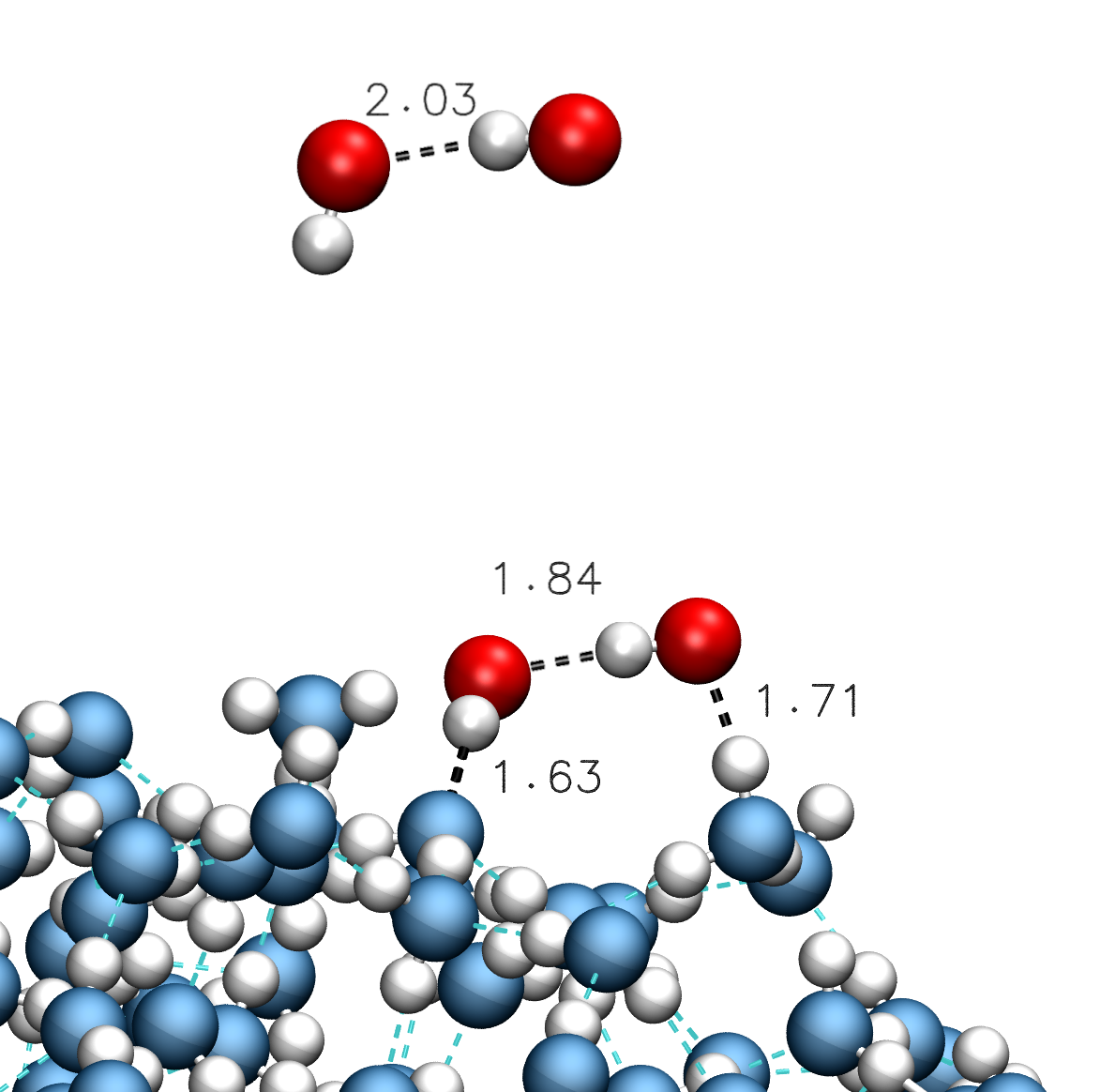}
  \caption{Gas phase (Top) and Surface (Bottom) minimal geometries showing the OH...OH hydrogen-bonded complex competitive with Reaction \ref{eq:central}. Minima were obtained at the CASPT2 level for the gas phase and with MLIP for large cluster.}
  \label{fig:singletOHOH}
\end{figure}

Our results clearly evince that inter-radical distance is an important factor for the probability of recombination, but it is not the only feature that defines whether radicals recombine or not. Because of that, additional structural features, such as radical orientation, were analyzed. A possible explanation for the minima in Figure \ref{fig:recomb_distance} can therefore be inferred from Figure \ref{fig:2Dplot} which shows endpoints of each trajectory in the 2D plot of distance between radicals (distance of oxygen atoms) and collective variable $\textrm{\textit{CV}}_{1}$, which is defined as:

\begin{equation}
    \textrm{CV}_{1} = \min \left( d_{\textrm{O}(1)\textrm{H}(2)}, d_{\textrm{O}(2)\textrm{H}(1)} \right),
\end{equation}
where $d_{\textrm{O}(1)\textrm{H}(2)}$ is the distance of oxygen from the first radical and hydrogen from the second radical, and $d_{\textrm{O}(2)\textrm{H}(1)}$ for oxygen from second radicals and hydrogen from first. Thus, $\textrm{\textit{CV}}_{1}$ represents the shortest OH distance of the non-bonded atoms in the studied radicals. From the position in this 2D plane, we can analyze not only the proximity of the radicals but also the orientation of H atoms on them. In the upper part of Figure \ref{fig:2Dplot} we see how in some of the trajectories that diffuse away, larger inter-radical distances are dominated by trajectories with higher added kinetic energy. This is obviously related to the ease of moving at larger distances for radicals with higher energy. In the bottom panel of Figure \ref{fig:2Dplot} we present a zoomed region of the top panel, focusing on the short and medium inter-radical ranges. At these ranges, we can identify two major clusters: one with the O-O distance around 1.4{~\AA} and $\textrm{\textit{CV}}_{1}$ of around 1.7{~\AA} which represents \ce{H2O2}; the other cluster shows O-O distance around 3{~\AA} and $\textrm{\textit{CV}}_{1}$ around 2{~\AA} and represents radicals forming of a hydrogen bond in the singlet state. Such clustering of final conformations in our simulations suggests the mechanism of the recombination and explains the minima present in Figure \ref{fig:recomb_distance} as we anticipated above. Visualizations of system geometry showing these conformations can be found in Figure \ref{fig:singletOHOH}. Hydrogen bonding apparently plays a role of competitive minimum on the potential energy surface (PES) in which radicals can be trapped at very low temperatures. Many trajectories with the initial radical distance around 3{~\AA} fell in this local minimum and stayed there until the end of the simulation, with not enough energy available to escape the minimum. The depth of this minimum was estimated to be about 0.21 eV (see Table \ref{tab:2}) at the CASPT2 level in the gas phase, without considering the effect of the environment. This minimum correlates well with what was found for trajectories with 0.5 and 1.0 eV of added kinetic energy, where the radicals have high enough energy to escape hydrogen bonding even though they visit the H-bonded minimum. It should be noted that radicals lose their energy as time progresses by dissipation into the water surface. 

As previously mentioned, the calculated FES shows a small barrier (of 0.05 eV) in the 2.5--3.0 \AA{} range. We cannot confidently indicate the origin of this barrier but is likely associated with the transition between the H-bond minima and the \ce{H2O2} potential well. Again, we remind that we present an averaged barrier, and therefore it is different from a barrier of a single reactive trajectory, which could happen both barrierlessly or with a barrier much higher than the one in Figure \ref{fig:PMF}. Finally, although we explained the minimum in recombination probability (Figure \ref{fig:recomb_distance}) for energies \{0.0, 0.1, 0.3\} as caused by the H-bond minimum, the large error bars make it difficult to provide a quantitative estimation of its depth. Despite the error bars, we are certain of the presence of a local minimum for recombination probabilities in Figure \ref{fig:recomb_distance}. The presence of the non-bonded dimer minimum in the CASPT2 gas phase calculations, summed to the presence of such minimum for multiple added energies and the visual inspection of the trajectories (Figure \ref{fig:singletOHOH}) indicate that the OH..OH minimum is the reason behind for such local minimum.  More trajectories would be necessary to quantitatively estimate its depth. However, it would require a number of trajectories beyond our current computational reach. Nevertheless, the disappearance of this feature at higher added energies reinforces our explanation.

\begin{table}[t]
\begin{center}
\caption{Depth of minima on PES for small complexes in the gas phase with respect to the separated radicals}
\label{tab:2}
\begin{tabular}{cc}
\toprule
Complex type & Energy [eV] \\
\bottomrule
Separated radicals & 0.0 \\
Hydrogen peroxide & 2.24 \\
Hydrogen bonded radicals & 0.21 \\
\bottomrule
\end{tabular}
\tablefoot{Energies were calculated for isolated complexes with CASPT2(6,4)/6-31+G(d,p) method. Every geometry was optimized on the same level of theory. Zero level was set as the energy of separated radicals.}
\end{center}
\end{table}

\subsubsection{The effect of the binding energy}

\begin{figure*}
  \centering
  \includegraphics[width=0.75\linewidth]{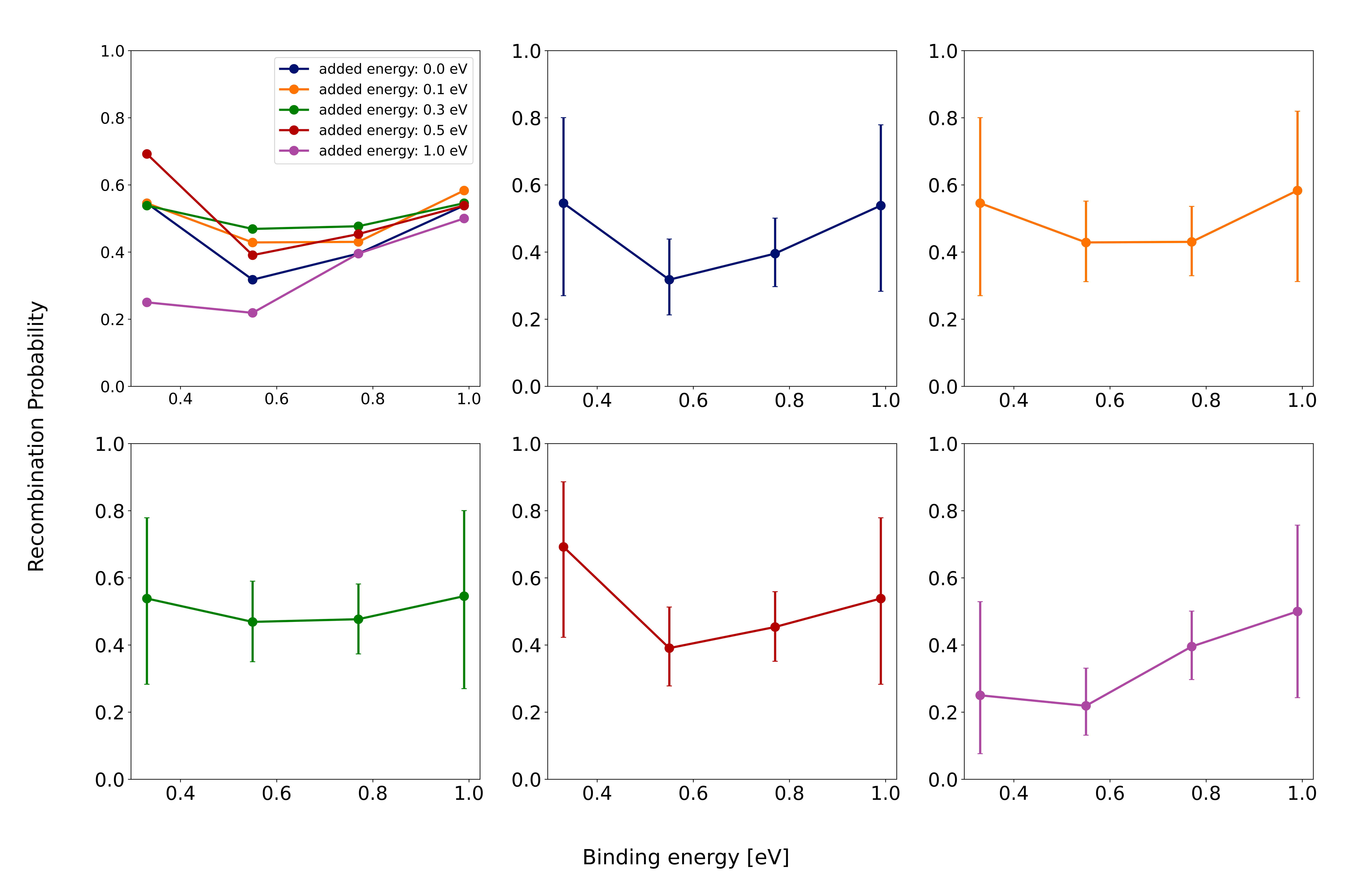}
  \caption{Recombination probability of hydroxyl radicals as a function of the binding energy of the OH radical. The binding energy of the more weakly bound radical is shown on the x-axis. The first top-left plot presents the results for all the energies combined. The following graphs present individually the datasets for different added energy, with the error bars included in the plots. Error bars represent Jeffreys Bayesian Interval generated by python library Statsmodels \citep{statsmodels}}
  \label{fig:recomb_BE}
\end{figure*}

We also explored the influence of the binding energy of OH in the recombination probability. We calculated binding energy as the difference between the energy of the whole cluster and a sum of cluster without radical and separated radicals. Therefore, the positive value of energy corresponds to spontaneous binding with the surface. It should be noted, that sampling of binding sites on the surface was affected by constraints during equilibration of the dynamics. As the main focus was on sampling the inter-radical distance. The sole presence of the second radical in proximity results in higher values of the binding energy in some cases, than what would be expected for a single OH on the water surface. Especially cases where the two OH are close together, e.g. binding energies close to 1.0 eV.  In general, we can conclude that due to these reasons, the presented binding energies are higher than the binding energy of one OH radical on a water surface. In previous studies of OH radical, the binding energy of a single hydroxyl radical were previously found to be 0.06-0.74 eV \citep{OH-Watanabe,BE_OH}. Figure \ref{fig:recomb_BE} shows the dependence of the recombination probability on the binding energy. As we said above, due to the interaction of OH radicals, higher binding energies correlate with lower distances of the two OH radicals, so the isolated effect of the binding site is difficult to capture. Due to the sampling process, the error bars in Figure \ref{fig:recomb_BE} are quite large for some values of binding energy. Detailed discussion of the effect of binding energy would therefore require a separate study. Nevertheless, we are able to see that there is no major effect of binding energy, as the graphs in Figure \ref{fig:recomb_BE} do not show any clear trend. In general, because, at least in this work, the binding energy serves as a proxy for the proximity of the radicals it can be said that the recombination probability depends mostly on the distance factor. This is also supported by the behavior for radicals with added kinetic energy, where there is an apparently lower recombination probability, in Figure \ref{fig:recomb_BE}. This is simple to rationalize, as radicals with higher kinetic energy are more likely to separate and escape the recombination, independent of the binding energy.

\subsection{Motion of radicals on ice surface} \label{sec:motion}

\begin{figure}
  \centering
  \includegraphics[width=0.60\linewidth]{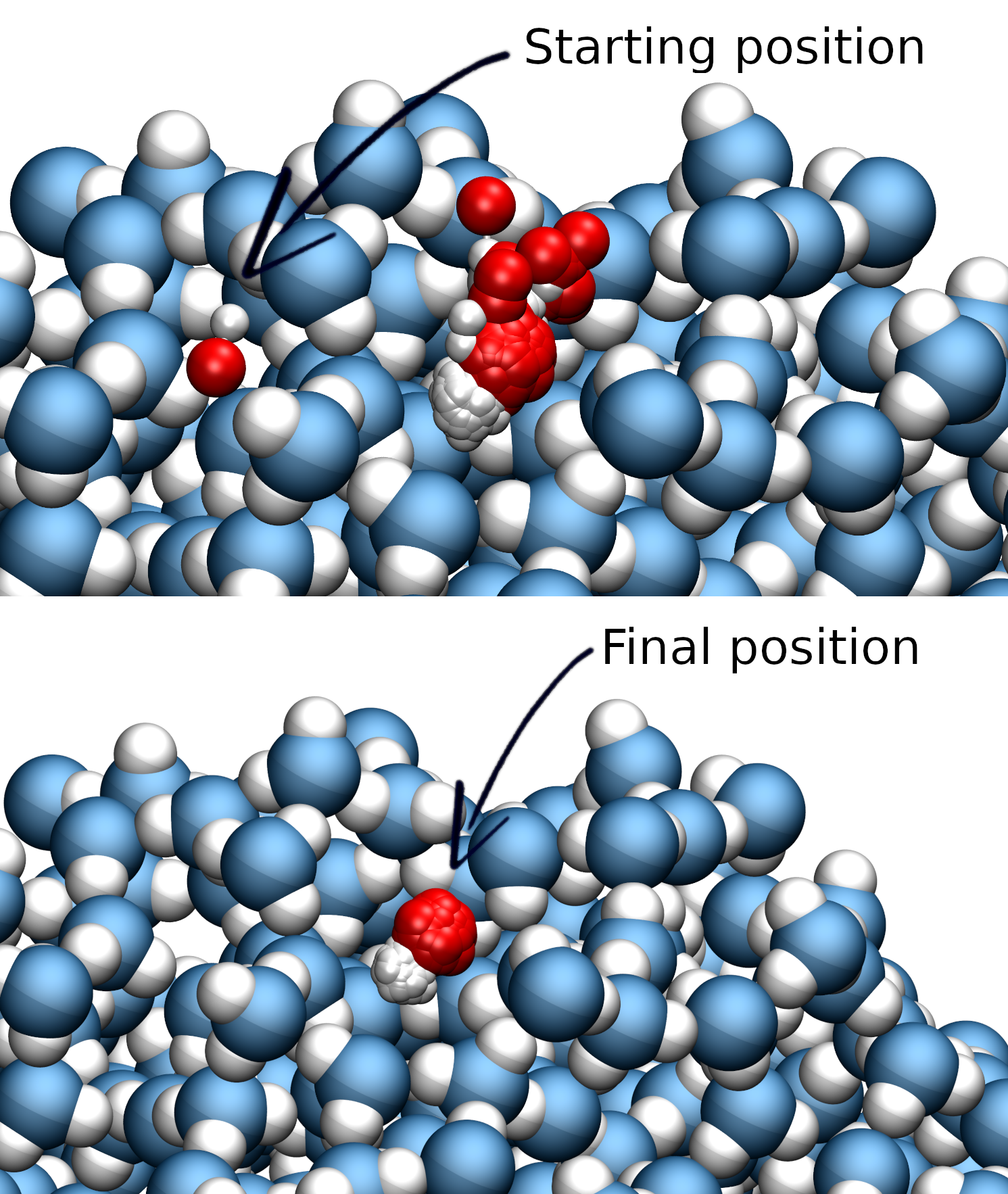}
  \caption{Visualization of dynamics of the OH radical on water surface after addition of 1.0 eV of additional translation energy. In the upper part of the picture, we can see all positions of the OH radical during the MD simulation. In the lower part, positions from the last nanosecond of the simulation are visualized. }
  \label{fig:motion}
\end{figure}

\begin{figure}
  \centering
  \includegraphics[width=0.75\linewidth]{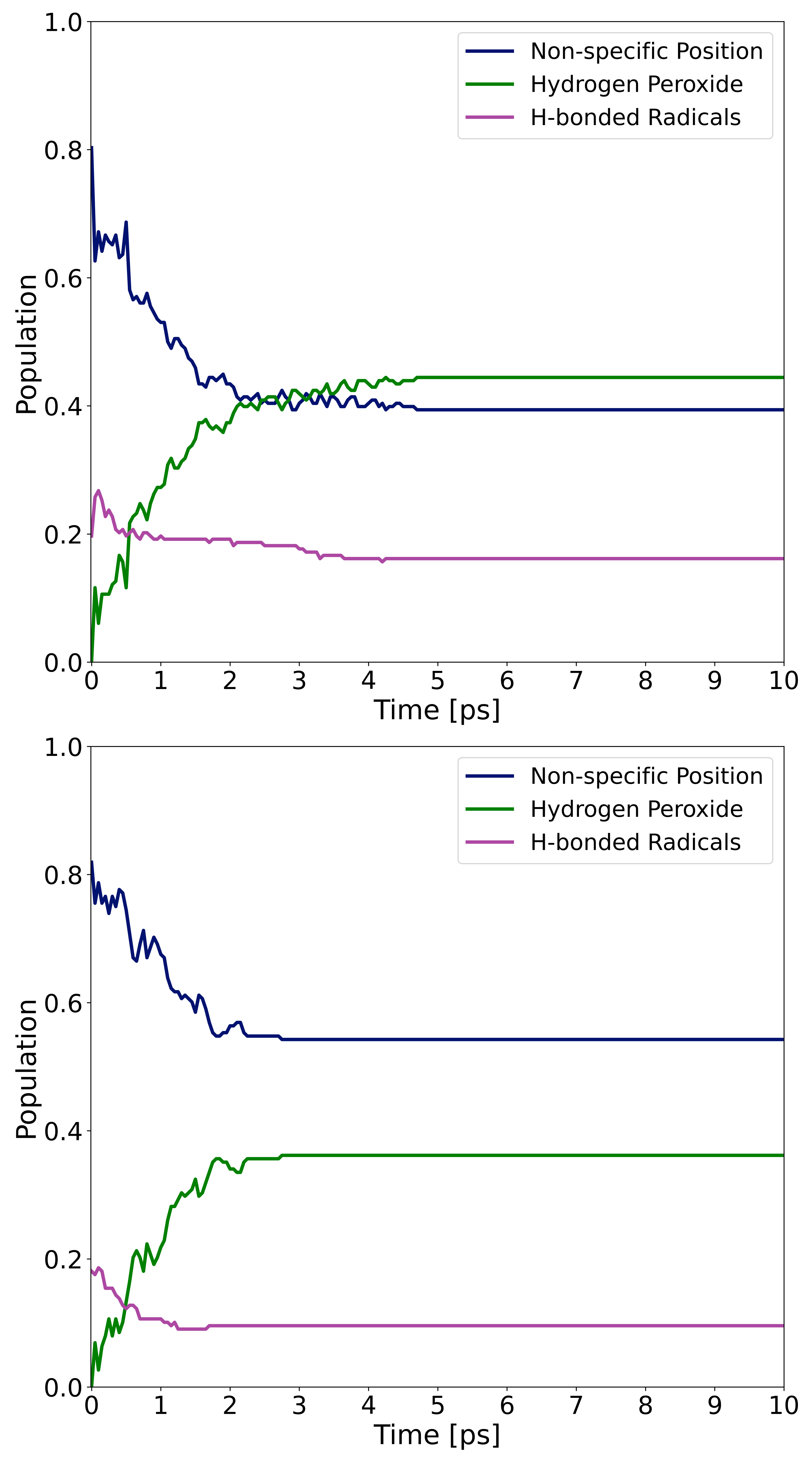}
x     \caption{Populations of H-bonding and hydrogen peroxide conformers in time for trajectories with 0.1 eV of added kinetic energy (upper plot) and 1.0 eV of added kinetic energy (lower plot).}
  \label{fig:conformers}
\end{figure}

\begin{figure}
  \centering
  \includegraphics[width=0.75\linewidth]{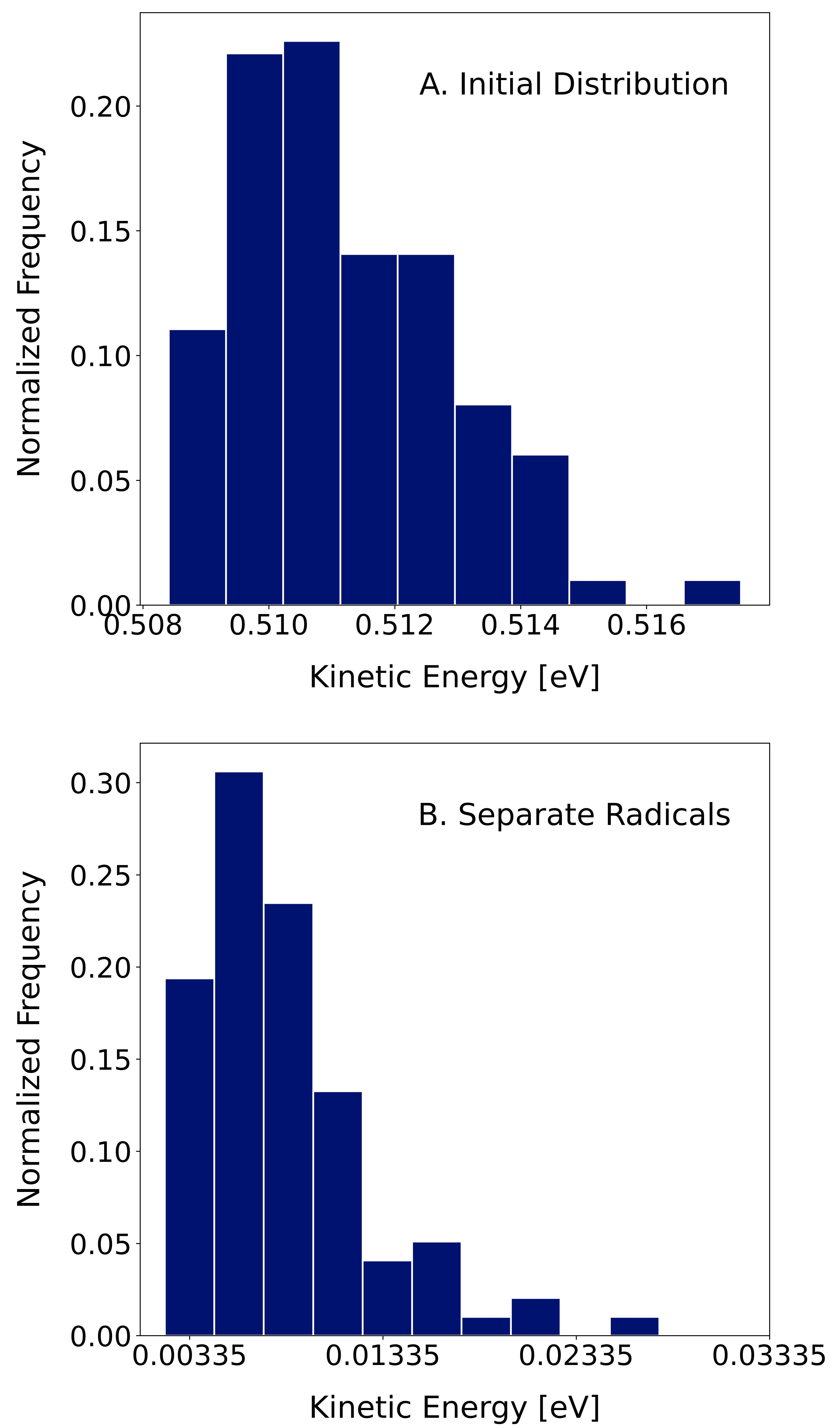}
  \caption{Distribution of kinetic energy for trajectories with 0.5 eV added. The upper plot shows kinetic energy for initial conditions, and the lower one describe averaged kinetic energy during the last 250 fs of MD trajectories.}
  \label{fig:kinetic_energy}
\end{figure}

For an even further understanding of the process, the populations of the conformers (OH...OH, \ce{H2O2}) in time are plotted in Figure \ref{fig:conformers}. The population dramatically changes during the first few ps. Higher kinetic energy of the radicals leads to faster stabilization of population levels. Extending the simulation further would not provide any qualitative change, given the low mobility of OH radical on cluster surface, which can be concluded both from the stabilization of populations in Figure \ref{fig:conformers} and from visualization in Figure \ref{fig:motion}. On the upper part of the visualization, we can see that at the beginning of the simulation, there is a transfer of the thermally excited radical from one position on the surface to another, but after the additional energy dissipates there is essentially no change in position and the radical stays in the same binding site (the lower part of Figure \ref{fig:motion}). In Figure  \ref{fig:conformers}, we observe differences in the behavior of low and high energetic radicals. For the added energy of 0.01 eV, we observe an increase in the population of H-bonded radicals and then a decrease of the population up until stabilization around 5 ps.  This phenomenon is also observed for other values of added energy, but as we see in the lower part of Figure \ref{fig:conformers}, is not present for the 1.0 eV. In this highly energetic case, the radicals are not as often trapped and the population of H-bonded radicals behaves similar to the population of radicals in non-specific position. Furthermore, as previously noted, separation is more effective for radicals with higher energy but as the additional translation energy will be dissipated to the water surface the radicals essentially freeze on the surface. This is supported by data in Figure \ref{fig:kinetic_energy} where distributions of kinetic energy at the start and at the end of the simulation are shown. We can see that even thermally excited radicals have very low kinetic energy at the end of the simulation. Therefore, radicals after 500 ps move very little regardless of the added kinetic energy at time zero.  

\section{Discussion, context of our results for astrochemical rate equation models} \label{sec:discussion}

\begin{table}[t]
\begin{center}
\caption{Recommended values of $\alpha$ for Reaction \ref{eq:central}.}
\label{tab:recommendations}
\begin{tabular}{cc}
\toprule
Regime & $\alpha$ \\
\bottomrule
Thermal & 0.33 $\pm$ 0.15 \\
Suprathermal (Radiolysis) & 1.00 \\
Suprathermal (Chemistry and Photons) & 0.40-0.66 \\
\bottomrule
\end{tabular}
\tablefoot{The value of 0.33 for thermal radicals is obtained from the second panel of Figure \ref{fig:recomb_distance}. The unity value for radiolysis is based on the lack of plateau at 3 \AA{} at high energies (Figure \ref{fig:recomb_distance}) indicating that non-reactive events come from diffusion out of the reaction center (not covered by the models). The values for chemical and photon energy come from the values at 3 \AA{} of the third, fourth, and fifth panels of Figure \ref{fig:recomb_distance}. The values collated in this table still do not include possible spin degeneracies, see last paragraph of Section \ref{sec:discussion}  }
\end{center}
\end{table}

Our results show that, despite the \textit{a priori} expectation that radical-radical coupling in Reaction \ref{eq:central} will have a unity probability of recombination, this is never the case in our simulations, even in the attractive region of the potential (free energy in our simulations). For \textit{thermalized radicals} (added energy = 0.0 eV) there are two reasons for this observation, one specific to the \ce{OH + OH} pair and another one that is general for any reaction on interstellar ices. Beginning with the latter, which is important at intermediate and long inter-radical distances, we attribute it to the presence of the water ice that emerges \textit{diffusion} barriers to an otherwise barrierless approach. This is essentially the same conclusion that, for example \citet{enrique-romero_quantum_2022} reached using static quantum chemical calculations. The specific reason as to why \ce{OH + OH} does not have a unity recombination probability is the formation of the OH...OH hydrogen bond complex, which precludes reaction. However, contrary to the first case where the barriers or lack thereof are a consequence of the interaction with the ice, the formation of the OH...OH complex is also found in the gas phase (See Figure \ref{fig:singletOHOH}), that is, it is not a consequence of the adsorbate ice interaction.

It is worth discussing what the presence of these punctual barriers and alternative minima entail for astrochemical models. Our reconstructed FES for the inter-radical distance shows that diffusion barriers are averaged out, making them absent or very small in the total reconstructed profile. This does not mean that they do not exist, our interpretation is that they simply are not kinetic barriers intrinsic to the \ce{OH + OH} reactive event. Thermal two-body reactions in astrochemical models follow a Langmuir-Hinshelwood formalism \citep{hasegawa_three-phase_1993}:
\begin{equation} \label{eq:thermal}
    k_{\textrm{t}} = \alpha \kappa_{ij} \left( \frac{1}{\tau\left(i\right)} \text{+} \frac{1}{\tau\left(j\right)} \right) \dfrac{1}{N_{\text{Site}}n_{\text{Dust}}},
\end{equation}
where $\alpha$ is the branching ratio of the reaction, $1/\tau(i) \text{+} 1/\tau(j)$ is the term dedicated to diffusion with $\tau(x)$ representing times for a single diffusion step and $1/{N_{\text{Site}}n_{\text{Dust}}}$ a term accounting for the number of binding sites ($N_{\text{Site}}$) and number density of dust grains ($n_{\text{Dust}}$). In reactions with competing processes like diffusion or desorption $\kappa_{ij}$ is obtained from the competition relation\citep{chang_gas-grain_2007}:

\begin{equation} \label{eq:comp}
   \kappa = \dfrac{k_{\rm reac}}{k_{\rm reac} + k_{\rm diff} + k_{\rm des}},
\end{equation}
with $k_{\rm reac}$, $k_{\rm diff}$ and $k_{\rm des}$ are the rate constants for reaction, diffusion and desorption (in the last two cases of the two possible reactants). In particular $k_{\rm reac}$ takes the form of:

\begin{equation} \label{eq:rb}
   k_{\rm reac} =  \nu \exp(-2(\gamma/\hbar)(2 \mu \beta)^{1/2}),
\end{equation}
with $\beta$, and $\gamma$ the variable parameters, and $\nu$ a term accounting for the characteristic frequency of the adsorbates, which in many models is simply the maximum of the characteristic frequencies of the two reactants and $\mu$ is the reduced mass. The two parameters have different meanings, with $\beta$ representing the reaction activation energy and $\gamma$ corresponds to the barrier width of the reaction, assuming a rectangular barrier for quantum tunneling. When $\beta$=0 (barrierless) $\kappa$ is made equal to unity (i.e. $\kappa$=1.0) effectively making the reaction depend solely on the diffusion terms and $\alpha$. In the context of the reaction studied in this work, it is reasonable to wonder what is the best way of including our recombination probabilities into models. There are two possibilities. First, we can vary $\alpha$ that could be taken directly from Figure \ref{fig:recomb_distance}. Second, we can vary $\beta$, whose value must be taken from the transition state of the \ce{OH + OH -> H2O2} reaction (not computed in this work). The latter approach can clearly lead to misleading results, at least in the context of Reaction \ref{eq:central} because it introduces an exponential term for every single reaction event, when our simulations clearly show that the probability of recombination is non-zero in Figure \ref{fig:recomb_distance} as it should be if an intrinsic barrier is present, because such a barrier could not be overcome in the short term of the molecular dynamics simulations. The approach of linearly weighting Equation \ref{eq:thermal} varying $\alpha$ is much more physically sound, at least at low temperatures up to 40 K where the \ce{OH} radical can slowly diffuse on ASW \citep{miyazaki_direct_2022}. 

However, it remains to be found out what should be value for $\alpha$ in these cases, from the different distances that we sampled in Figure \ref{fig:recomb_distance}. With this in mind, it is clear that the recombination rate does depend on the origin of the radicals not only through their initial energy but also through their starting position, which is a key factor to take into account when setting the value of $\alpha$. We think that the values of 3 \AA, just before the effective barrier (See Figure \ref{fig:PMF}) are an excellent compromise to not double-count diffusion. Thus, $\alpha$ for Reaction \ref{eq:central} in a thermal regime is 0.33 (also gathered in Table \ref{tab:recommendations}). When temperature increases, it is more difficult to recommend a particular scheme, as emerging barriers could be overcome thermally. In those cases the best compromise is likely to set $\alpha$=1.0 as normally done. At this stage of research, the arguments and recommendations given in this paragraph can only be applied to Equation \ref{eq:central}. Although we think they must be general to any interstellar grain reaction on ASW having both barrierless and non-barrierless channels depending on the orientation, we do not have a sufficiently big body of evidence to confirm it, so we postpone the study of more reactions in future works.

The second case to be discussed is the case of suprathermal OH whose rate constant does not have any diffusion terms. While several formalisms are available \citep{shingledecker_cosmic-ray-driven_2018, Jin2020} we can generally consider that all reactions are effectively barrierless, making the rate constant essentially proportional to the branching ratio, $\alpha$, $k_{\textrm{t}} \propto \alpha$. We note that in reality, this is only true if the amount of energy is enough to submerge otherwise emerged barriers, as we recently discussed in \citet{del_valle_formation_2024}. However, the recombination probability derived for suprathermal OH is not unique, as it depends on the added energy of the OH radical (Figure \ref{fig:recomb_distance} and \ref{fig:recomb_BE}). The inclusion of this heterogeneity into models is complex as it requires treating species differently depending on the added energy, which is a continuous function. Alternatively, an average value could serve as a compromise, assuming the distribution of translational energies upon excitation follows a certain distribution. This latter solution also has problems, since different excitation mechanisms, chemical, photons, or cosmic rays, do not have similar energy distributions. Moreover, our sampled cases top at an added energy of 1.0 eV, which is reasonable for simulating a fraction of the energy deposited by photons but a lower bound for cosmic rays initiated chemistry \citep{shingledecker_cosmic-ray-driven_2018}, for example, and an upper bound for chemically activated reactions. Despite the uncertainties, we think that a value of $\alpha$=1.0 for radiolysis is reasonable. We rationalize this choice on the basis of the lack of plateau in the rightmost bottom panel of Figure \ref{fig:recomb_distance}, meaning that the H-bonded complex will not be formed and the lower recombination probability is a consequence of diffusion out of the reaction site, something already considered in the models and not intrinsic to the reaction event. For lower energy processes a value between 0.40 and 0.62 could be used, based on the same observations that, at intermediate added energies, the plateau in recombination probabilities still appears, indicating the formation of a H-bond complex. As in the case of the thermal case, the study of more systems with this methodology will allow us to determine whether the case of Reaction \ref{eq:central} can be extrapolated to any suprathermal reaction in the ISM, although we anticipate that in the case of Reaction \ref{eq:central}, the presence of the vdW complex shown in Figure \ref{fig:singletOHOH} makes this a particular case.

Lastly, we discuss the effect of the different spin states in astrochemical models as the \ce{OH + OH} reactants can also appear in the triplet state, as stated in the Introduction \citep{redondo_reconstruction_2020}. To properly weigh the contribution of the different spin states of a reaction one should weigh the branching ratios. The population of the relative spin states of two interacting radicals can, however, strongly change in time. If we analyze the gas phase case, the weighting factor for a reaction can be obtained from the ratio of spin multiplicities of each channel ($M$). For the reaction studied in this work, \ce{OH + OH -> H2O2}, $M_{\rm 1}$ = 1 whereas for the competing channel, \ce{OH + OH -> H2O + $^{3}$O}, $M_{\rm 2}$ = 3. Thus, the weighting factor for \ce{OH + OH -> H2O2} is therefore 1 / (1 + 3) = 0.25 and for \ce{OH + OH -> H2O + $^{3}$O} it is 0.75. Looking now at the reactions on the surface, the non-diffusive reaction should essentially behave as in the case of the gas phase, because the timescale for the reaction (around picoseconds), see, for example, \citep{shingledecker_cosmic-ray-driven_2018}) is probably too low for intersystem crossing to occur. In the case of a diffusive reaction (thermal regime), the situation can be much more complicated. At 10 K, with the diffusion of OH being effectively null, and the rate for any reaction will be insignificant. At temperatures close to the diffusion temperature of the OH radical, 36~K \citep{miyazaki_direct_2022}, the situation is much more complex. In singlet state the radicals would move on potential of mean force mixed given by a Boltzmann weighting of the multiple potentials. This potential would effectively be very similar to the lowest energy potential as done in our simulations. The possibility of intersystem crossing might not be negligible for longer timescales, thus enhancing the total recombination rate. Our recommendation is, in the absence of non-adiabatic simulations (beyond the scope of this work), keep using the quotient of spin multiplicities, but recognize the potential source of uncertainty. However, we advocate for an increasingly more complex treatment of different spin channels in surface astrochemistry. Finally, these correction factors can be incorporated in models by multiplying $\alpha$, such as those collated in Table \ref{tab:recommendations}.

\section{Conclusions} \label{sec:discussion:conclusions}

The study explored the recombination dynamics of hydroxyl OH radicals on amorphous solid water (ASW) surfaces using molecular dynamics simulations driven by machine-learned interatomic potentials (MLIP). This reaction had not been theoretically described on water surfaces before, largely due to significant methodological challenges that were overcome in this work through the use of novel computational techniques. As far as we know, this is the first study to employ neural networks trained on hybrid multi-reference/DFT data for molecular dynamics. Although multi-reference data labeling requires extensive data points and specialized expertise, it emerged as the best compromise between accuracy and feasibility for simulating radical reactions. The methodology presented in this study sets the foundation for future non-adiabatic simulations in large systems. 

Our findings describe the mechanism of hydroxyl radical recombination and highlight the significance of the local minimum of hydrogen-bonded radicals, which plays a crucial role as a competitive product for thermal OH radicals. At low temperatures, this hydrogen-bonding minimum reduces the probability of recombination for radicals starting at a distance of larger than 3 \AA. This effect is only partially overcome when additional energy is introduced to the OH radicals. Radicals with higher initial kinetic energy were also less likely to recombine due to their greater ability to separate and move across the surface. Regarding the influence of binding energy, our study demonstrates that the initial inter-radical distance has a greater effect on recombination probability than the binding energy itself. However, binding energies and inter-radical distances are correlated and weighting specific contributions is difficult.

We consider the recombination reaction essentially barrierless, with the barrier being or not being present based on the character of the surface and the initial motion of the radicals on it. In the total reconstructed profile, such diffusion barriers are averaged out. Based on these results, we recommend adjusting the branching ratio parameter (\(\alpha\)) in Equation \ref{eq:thermal} in astrochemical models for both thermal and suprathermal reactions. For thermal reactions, a value of \(\alpha \approx 0.33\) is suggested, while for suprathermal processes, a value closer to 1.0 is more appropriate, topping $\alpha$ at 1.0 for radiolytic processes. These recommendations pave the way for more refined astrochemical models that can better simulate the complex chemistry occurring on interstellar dust grains. 

\begin{acknowledgements}

J.P. and P.S. were supported by the Czech Science Foundation (EXPRO project no. 21-26601X), by the grant of Specific university research – No. A2\_FCHI\_2023\_011 from UCT Prague and supported by the Ministry of Education, Youth and Sports of the Czech Republic through the e-INFRA CZ (ID:90254). J.P. is a student of the International Max Planck Research School “Quantum Dynamics and Control”.  G.M acknowledges the support of the grant RYC2022-035442-I funded by MCIU/AEI/10.13039/501100011033 and ESF+. G.M. also received support from project 20245AT016 (Proyectos Intramurales CSIC). We acknowledge the computational resources provided by bwHPC and the German Research Foundation (DFG) through grant no INST 40/575-1 FUGG (JUSTUS 2 cluster), the DRAGO computer cluster managed by SGAI-CSIC, and the Galician Supercomputing Center (CESGA). The supercomputer FinisTerrae III and its permanent data storage system have been funded by the Spanish Ministry of Science and Innovation, the Galician Government and the European Regional Development Fund (ERDF).

\end{acknowledgements}

%
%

\bibliographystyle{aa}
\bibliography{references.bib}

\begin{appendix}

\section{Consistency tests between the machine learned potential and the reference \textit{ab initio} method.} \label{sec:apA}

\begin{figure}
  \centering
  \includegraphics[width=0.75\linewidth]{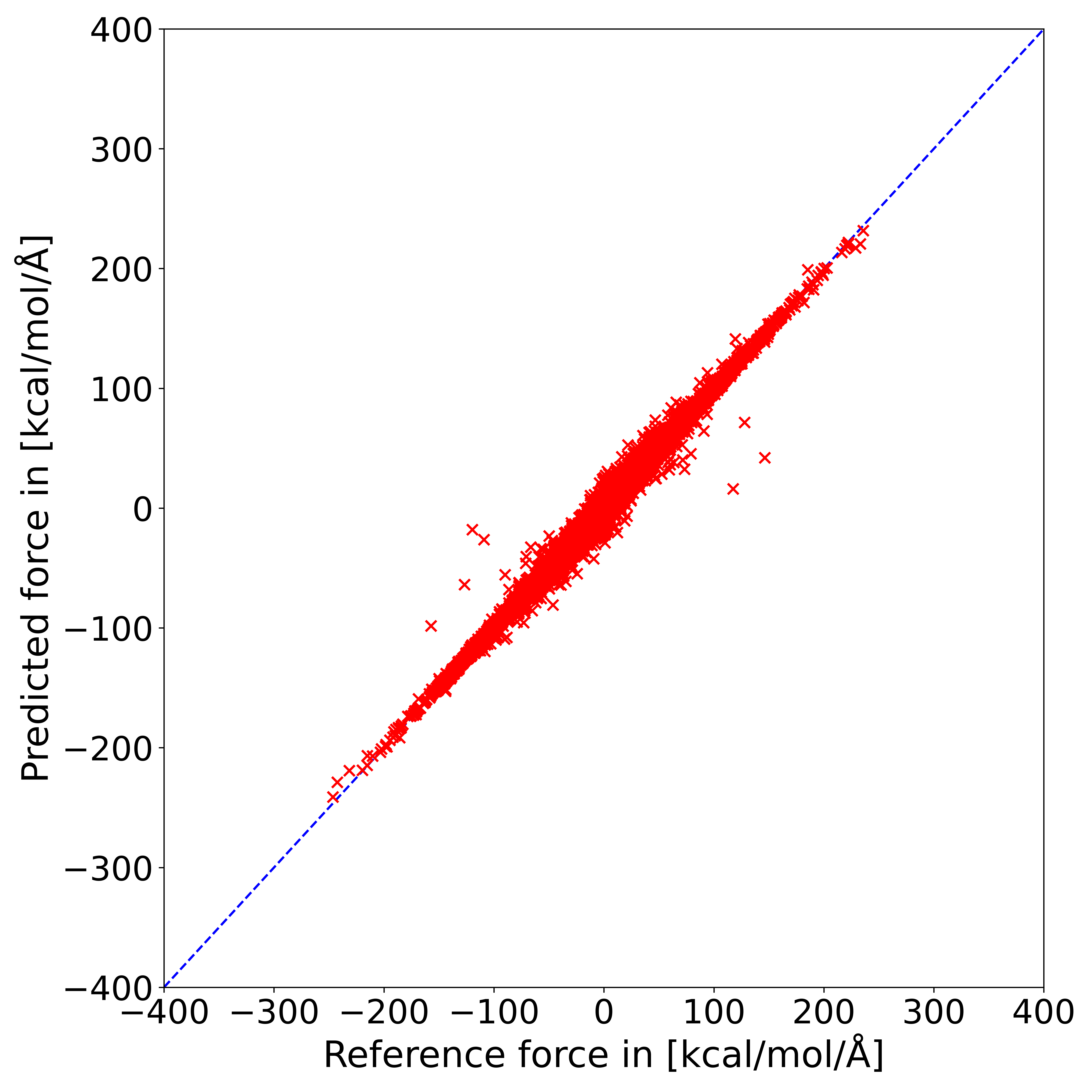}
  \caption{Analysis of reference vs predicted forces for geometries from test set (part of original data set which was previously unseen by the model). The blue diagonal line is the function y = x and therefore represents perfect agreement of reference and prediction.}
  \label{fig:test_set}
\end{figure}

To check the quality of MLIP used for the simulations, we employed several tests. Firstly, we determined the mean absolute error (MAE) and root-mean-square error (RMSE) on the data in our test set. The test data were sampled from the original data set same as the training and validation set but were never seen by the model during the training phase. The MEA for forces was 0.44 kcal/mol/\AA{} and the RMSE was 0.83 kcal/mol/\AA. These values are averaged over all 3 models, which were trained simultaneously, their average was also used for all production calculations. Figure \ref{fig:test_set} shows the performance of the calculated forces against the reference force computed by QM/QM(CASPT2(6,4/6-31+G(d,p) + PBE(D3BJ)/def2-SVP). Apart from few outliers, the data lie on the diagonal and therefore confirm the quality and usability of our model. Outlying geometries can be predicted during the production runs with uncertainty computed from the ensemble of models.

\begin{figure}
  \centering
  \includegraphics[width=0.75\linewidth]{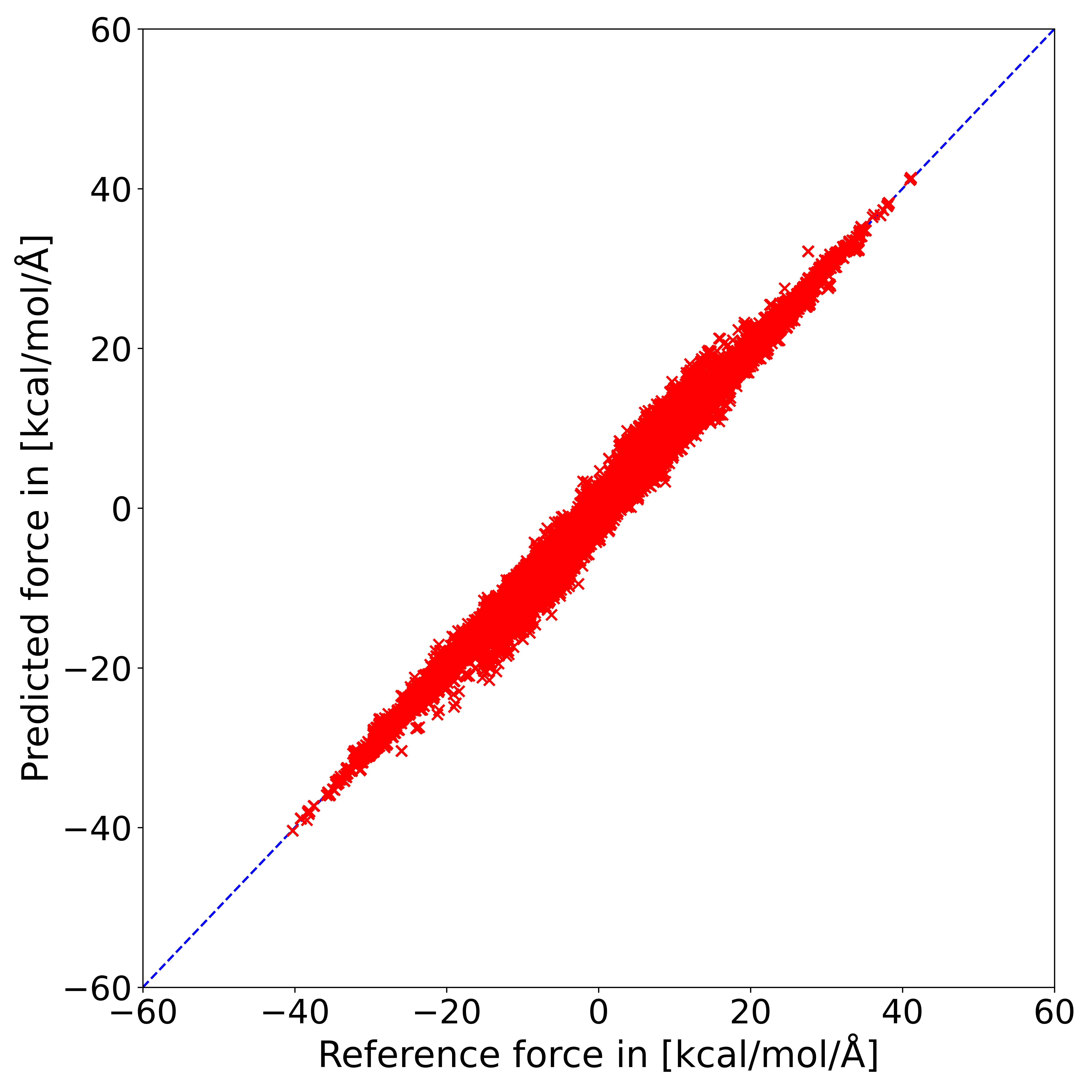}
  \caption{Analysis of reference vs predicted forces for geometries produced by molecular dynamics of OH radical impacting on water cluster. The blue diagonal line is the function y = x and therefore represents perfect agreement of reference and prediction.}
  \label{fig:impact}
\end{figure}

An important process in simulations on cluster surfaces is the evaporation of molecules from the cluster surface. To test our model with respect to this process, we produced trajectories with one radical impacting a cluster surface. Such geometries were added to the data set for small clusters. To estimate the performance in this manner, we present Figure \ref{fig:impact} in which reference vs predicted forces are presented for a data set of geometries from impacting trajectories similar to those in the original data set. In principle, such trajectories represent clusters with one of the radicals at different distances from the cluster surface up the cutoff 5.5 \AA{} where no interaction with the surface is present. With all data points lying on the diagonal, we concluded that our model is performing well with regard to interaction with the surface.

\begin{figure}
  \centering
  \includegraphics[width=0.75\linewidth]{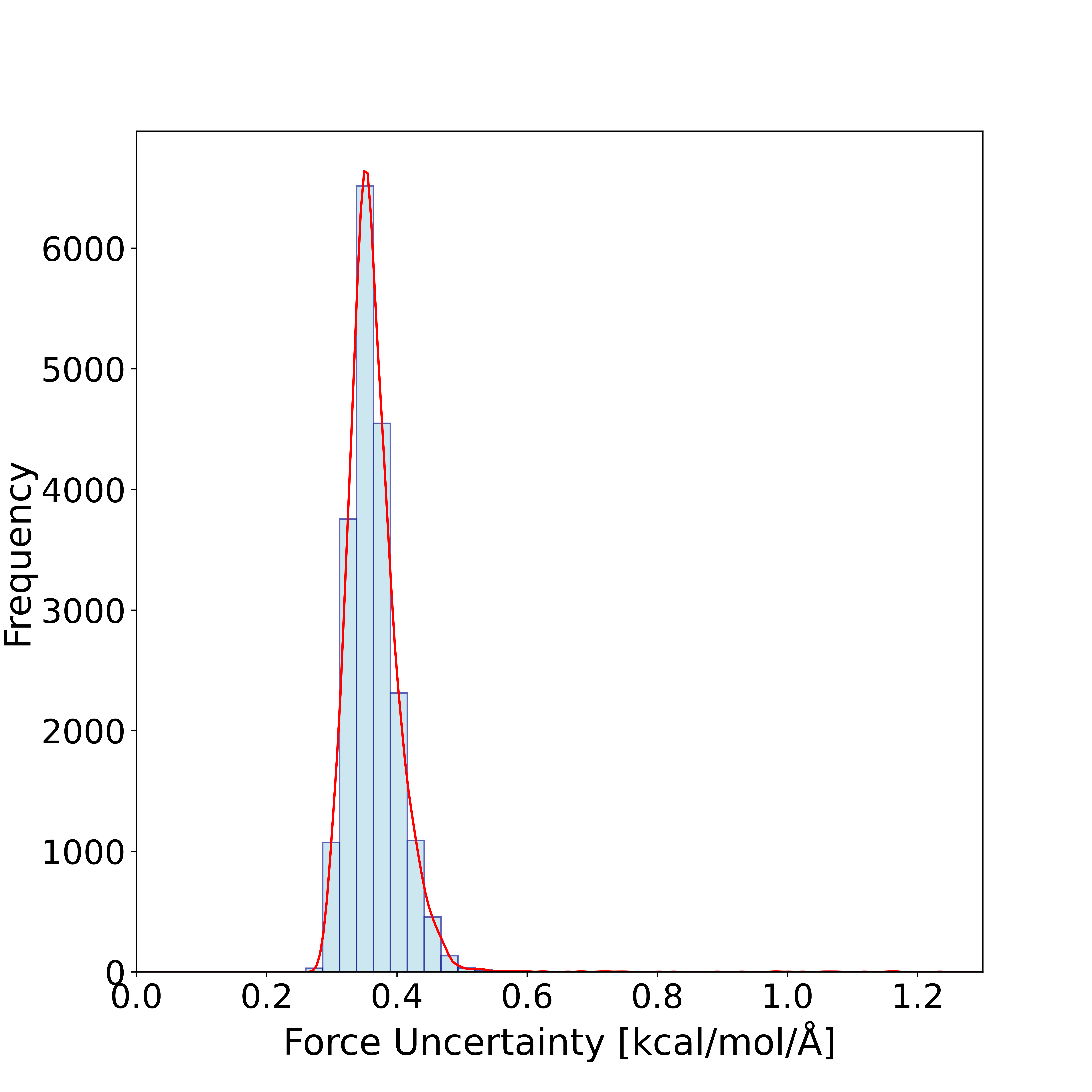}
  \caption{Force uncertainty tracked along the MD trajectory of relaxed molecular dynamics of radicals moving on the cluster surface. Uncertainty was computed from the ensemble of 3 models and was tracked during the whole time of the simulation.}
  \label{fig:md_force}
\end{figure}

The uncertainty of force prediction was tracked along the MD simulations to avoid artificial results coming from possible prediction failure. In Figure \ref{fig:md_force} force uncertainty during NVE MD simulation is presented. The model performs very well, with the force uncertainty not exceeding 0.6 kcal/mol/\AA. Such performance is way below the threshold of 1 kcal/mol/\AA{} which is often considered a threshold for chemical accuracy. 

The produced models performed well in all proposed tests and were therefore used for production calculations. We also note, that the fact of training multiple models allowed for checking the performance in any further calculations.

\section{Supplementary analysis of performed MD trajectories.} \label{sec:apB}

\begin{figure}
  \centering
  \includegraphics[width=0.75\linewidth]{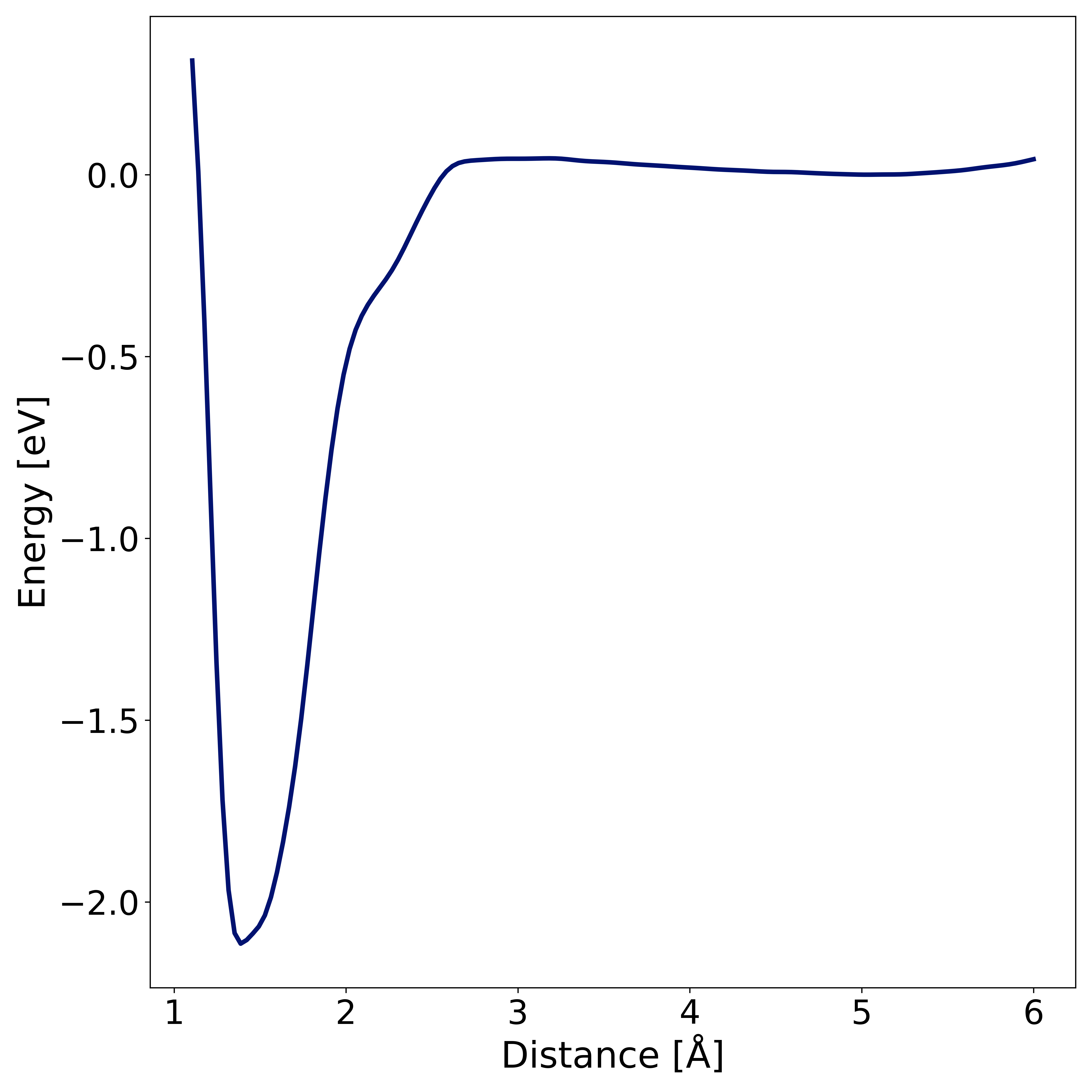}
  \caption{Potential of mean force from metadynamics using smaller (0.001 eV) height of Gaussian functions.}
  \label{fig:small_pmf}
\end{figure}

To address expected questions and provide additional support for our data, we present supplementary graphics analyzing the MD trajectories central to this study. In Figure \ref{fig:small_pmf}, we present graphics similar to those in Figure \ref{fig:PMF} from the results section. In this case, a Gaussian function with a height 10 times smaller was used to sample the observed barrier more cautiously. The presence of the barrier was also confirmed in this calculation; however, fully converging this simulation with higher accuracy would require an impractical amount of computational time. Therefore, we focus on presenting Figure \ref{fig:PMF} in the results section, as it provides more robust conclusions.

\end{appendix}

\end{document}